\documentclass[12pt]{article}
\usepackage{amsmath}
\pdfoutput=1
\usepackage{enumerate}
\usepackage{natbib}

\usepackage[english]{babel}

\usepackage{latexsym}
\usepackage{amssymb}
\usepackage{amsfonts}
\usepackage{bm}
\usepackage{graphicx,epsfig}
\usepackage{lscape,amsmath,amsxtra}
\usepackage{amsthm}
\usepackage{eucal,mathrsfs}
\usepackage{color}
\usepackage{srcltx}
\usepackage{rotating}

 \newtheorem{cor}{Corollary}

\def\E{\mathrm{E}}
\def\var{\mathrm{var}}
\def\cov{\mathrm{cov}}

\def\query[#1]{\fbox{\texttt{#1}}}

\newcommand{\blind}{0}

\begin{document}

\def\spacingset#1{\renewcommand{\baselinestretch}%
{#1}\small\normalsize} \spacingset{1}


\if0\blind
{
  \title{\bf Comparing composite likelihood methods based on pairs for spatial Gaussian random fields}
  \author{Moreno Bevilacqua\\
    DEUV, Universidad de Valparaiso - Valparaiso, Chile\\
    and \\
    Carlo Gaetan\\
    DAIS, Universit\`a  Ca' Foscari - Venezia, Italy.}
  \maketitle
} \fi

\if1\blind
{
  \bigskip
  \bigskip
  \bigskip
  \begin{center}
    {\LARGE\bf Title}
\end{center}
  \medskip
} \fi

\bigskip
\begin{abstract}
In the last years there has been a growing interest
in proposing   methods for estimating covariance functions
for geostatistical data.
Among these,  maximum likelihood estimators have  nice features  when we deal with
a Gaussian model.  However maximum likelihood becomes impractical when the number of observations is very large. In this work we review some solutions  and we contrast them in terms of loss of statistical efficiency and computational burden.
Specifically we focus on three types of weighted composite likelihood functions based on pairs and we compare them with the  method of covariance tapering. Asymptotics properties of the three estimation methods  are derived.
We illustrate the effectiveness of the methods through theoretical examples, simulation experiments
and by analysing a data set on yearly total
precipitation anomalies at weather stations in the United States.
\end{abstract}

\noindent%
\vfill

\newpage
\spacingset{1.45} 
\section{Introduction}
The geostatistical approach models  
data coming from a limited number of monitoring stations
 as a partial realization from a 
spatial random field  defined on the continuum space.
The literature persistently emphasises the importance of the estimation of the covariance
function  for several reasons. For instance, the best linear
unbiased prediction  at an
unobserved site  depends on the knowledge of the
covariance function of the process  \citep{Cressie:1993}.
Since a  covariance function must be positive definite,
 practical estimation generally requires the selection of
some parametric classes of covariances and  the corresponding  estimation
of these parameters.

The maximum likelihood method  is  generally
considered the best method for estimating the parameters of
covariance models. However, for a Gaussian random field with a given parametric covariance
function, exact computation of the likelihood requires calculation of the
inverse and determinant of the covariance matrix, and this evaluation  is
slow when the number of observations is large.

More precisely, let $\{Z(s),  s \in \mathbb{R}^d\}$ be a stationary Gaussian random field
with  zero mean  and covariance function $\cov(Z(s),Z(s'))=\sigma^2\rho(s-s';\phi)$,
 $s,s'\in \mathbb{R}^d$, \textbf{$\sigma^2>0$, $\phi \in \Phi \subset R^p$}. The unknown  parameters $\theta=(\sigma^2,\phi)^\intercal $ must be estimated on the basis
of a finite number of $n$  observations $Z=(Z(s_1),\ldots,Z(s_n))^\intercal $.	

  The
 log-likelihood function for a
 the Gaussian
random field can be written as
\begin{equation}\label{eq:loglik}
l(\theta)=-\frac 12\log|\Sigma(\theta) |-\frac 12{Z}\,^\intercal [\Sigma(\theta)]
^{-1}{Z},
\end{equation}
where $[\Sigma(\theta)]_{ij}=\sigma^2\rho(s_i-s_j;\phi)$.

The
most critical part when evaluating (\ref{eq:loglik})   is to evaluate the
determinant and inverse of $\Sigma(\theta)$. This evaluation could
be theoretically carried out in $O(n^{2.81})$ steps (see, e.g.,
\citet{Aho:Hopcroft:Ullmann:1974}, although the most widely used
algorithms such as Cholesky decomposition require up to $O(n^ 3)$
steps. This can be prohibitive if $n$ is large. This motivated
to look for either approximations to the likelihood function
or different minimum-contrast-type methods that require less
than $O(n^3)$ steps to evaluate \citep{Whittle:1954,Vecchia:1988,Curriero:Lele:1999,Stein:Chi:Welty:2004,Caragea:Smith:2006,Fuentes:2007,Kaufman:Schervish:Nychka:2008,Cressie:Johannesson:2008,Stein:2008,
Lindgren:Rue:Lindstrom:2011}.

Significant computational gain is achieved when the sampling
locations form a regular lattice. In this case, the covariance
matrix has a special structure \citep{Whittle:1954} that can be exploited by using
spectral methods, reducing the computational burden. For irregularly
spaced data,  \citet{Fuentes:2007} extended the Whittle's idea and
suggested to integrate the spatial process over grid cells,
obtaining an approximation to the likelihood  on a lattice
structure.
The method requires $O(n \log_2 n)$ operations and does not involve calculating
determinants.

 \citet{Rue:Tjelmeland:2002} approximated the
inverse of covariance matrix to be the precision matrix of a
Gaussian Markov random field wrapped on a torus.
In this case the numerical factorization
of the precision matrix  can be done at a cost
of $O(n^{3/2})$ for a two-dimensional Gaussian Markov random field.
Recently \citet{Lindgren:Rue:Lindstrom:2011} exploited  the representations of certain Gaussian
random fields with Mat\'ern covariance structure by the  solution of a stochastic partial differential equation and derived an approximation
based on a Markov Gaussian  random field  with sparse precision matrix.
The main drawback of this approach is that we can only find the explicit form  for those
Gaussian random fields that have a Mat\`ern covariance structure at certain degree of
 smoothness \citep[see the discussion to ][]{Lindgren:Rue:Lindstrom:2011}.

Another idea \citep{Banerjee-Gelfand-Finley-Sang:2008,Cressie:Johannesson:2008,Stein:2008} is  putting a low rank structure on the covariance matrix, namely
\begin{eqnarray*}
 \Sigma(\theta)&=& S K(\theta_0)S^\intercal  + \theta_1 V
\end{eqnarray*}
where   $S$ is a known $n \times r$ matrix,  $K(\theta_0)$  a $r\times r$ positive definite matrix depending on the unknown  parameter $\theta_0$,
$V$  a known diagonal matrix,  $\theta_1 \ge 0$ an unknown scalar, and $\theta = (\theta_0 , \theta_1 )$.
This allows  to calculate the inverse and the determinant of a large covariance by inverting and calculating the determinant of a matrix of lower dimension ( $r \ll n$ )
according to the Woodbury formula \citep[see,][for instance]{Cressie:Johannesson:2008}.

All these methods have their relative strengths  but they can lead to making unnatural
assumptions about the random fields giving a less appropriate model.
Instead in the sequel we  will concentrate on two estimation methods  that  preserve the starting model,
and, with some adjustments, allow us   to perform standard inference as in the case of classical likelihood estimation.


In the tapering approach \citep{Kaufman:Schervish:Nychka:2008} certain elements of the
covariance matrix  that correspond to pairs with large distance  are set to zero.
This is done, see Section \ref{sec:taper}, in a way to preserve the property of positive definitess in the resulting 'tapered' matrix.
Then sparse matrix algorithms  can be
used to evaluate efficiently an approximate likelihood where the original covariance has been replaced by the 'tapered' matrix.
The intuition behind this approach is that correlations between
pairs of distant sampling locations are often nearly zero, so little
information is lost in taking them to be independent.

With composite likelihood  (CL) we will indicate a general class of estimating functions  based on the
  likelihood of marginal or conditional events (\citet{Lindsay:1988}, \citet{Varin:Reid:Firth:2011}).
 This kind of estimation method can be  helpful when  it is difficult to evaluate or to specify the full likelihood.
  In our case  the evaluation of the  likelihood of the whole set of the observations is too expensive and composing likelihoods with a smaller number of observations is  computationally appealing.

 Different types of CL functions have been proposed in literature for estimating the covariance model of spatial and spatio-temporal Gaussian random fields. For instance \citet{Stein:Chi:Welty:2004} proposed a  CL based on conditional events improving a previous proposal of \citet{Vecchia:1988}. More recently,
    \citet{Bevilacqua:Gaetan:Mateu:Porcu:2012} considered a weighted CL based on difference of Gaussian pairs in the space time context and  \citet{Eidsvik:Shaby:Reich:Wheeler:Niemi:2013} developed a pairwise Gaussian block composite likelihood in the same vein of \citet{Caragea:Smith:2006}.

As outlined in \citet{Lindsay:Yi:Sun:2011}, for a given estimation problem the choice of a suitable  CL function  should be driven by statistical and computational considerations.
In particular, for Gaussian random fields, there is a clear computational advantage
when we consider only  CL based on pairs of observations.

Therefore in this paper
we contrast CL functions based on the marginal distribution of a pair or
the  distribution of an observation conditionally to another observation or
the distribution of the difference between two observations.
Since the  three CL functions are equivalent from computational point of view, the main purpose of the paper is to compare them from statistical efficiency point of view.
Moreover we  establish the asymptotic properties of the associated estimators.
Lastly we  argue that the CL approach based on  pairs is a valuable competitor of the tapering approach with respect to the efficiency when the computational burden is heavy.
This is done through theoretical examples and  simulations.

The paper is organized as follows.
 In Section \ref{sec:taper} we present in more detail  the tapering method
 while  Section \ref{sec:CL} describes the three CL estimating methods based on Gaussian pairs.
In Section  \ref{sec:numerical_examples} we compare the methods described in Section  \ref{sec:taper}  and \ref{sec:CL} through theoretical examples and numerical results.
As a real data example, in Section \ref{sec:dataset}  we apply CL and tapering methods  on  a real data set of yearly total precipitation anomalies already analyzed in
 \citet{Kaufman:Schervish:Nychka:2008}.
Finally, in Section \ref{sec:conclusions} we give some conclusions.

\section{Tapered likelihood}\label{sec:taper}

In the tapering approach, pionered by \citet{Kaufman:Schervish:Nychka:2008}, certain elements of the
covariance matrix  $\Sigma(\theta)$ are set to zero multiplying
$\Sigma(\theta)$ element by element by a correlation matrix coming
from a compactly supported isotropic correlation function.
More precisely,  we consider  a correlation function
$r(s-s';d)$ that is identically 0 whenever $\|s-s'\| \ge d >0$.
The `tapered'
matrix
$\Sigma_T(\theta)= \Sigma(\theta) \circ R(d),
$
where $[R(d)]_{ij}=r(s_i-s_j;d)$ and $\circ$ is the Schur product,
is still positive definite and
 sparse matrix algorithms  can
be used to evaluate an approximated log-likelihood efficiently  \citep{Furrer:Sain:2010}.
There are several ways to construct compactly supported correlation function
\citep{Gneiting:2002b} and an example that we consider in the sequel is given by a specific type of  Wendland
function
\begin{equation}\label{eq:wendland}
  r(h;d)=\left(1-\frac{\|h\|}{d}\right)^4_+\left(1+4\frac{\|h\|}{d}\right)
\end{equation}
with $(a)_+=\max\{0,a\}$.
Our choice is supported by the asymptotic results in \citet{Kaufman:Schervish:Nychka:2008} and   \citet{Du:Zhang:Mandrekar:2009}.

 \citet{Kaufman:Schervish:Nychka:2008}
  proposed two approximations of the log-likelihood (\ref{eq:loglik}), namely
\begin{equation}\label{eq:tap1}
l_{{T,1}}(\theta,d)=-\frac 12\log|\Sigma_T(\theta) |-\frac 12{ Z}\,^\intercal [
\Sigma_T(\theta)]
^{-1}Z,
\end{equation}
and
\begin{equation}\label{eq:tap2}
l_{{T}}(\theta,d)=-\frac 12\log|\Sigma_T(\theta) |-\frac 12{Z}\,^\intercal ([\Sigma_T(\theta)]^{-1}\circ R(d))Z
.
\end{equation}
In  (\ref{eq:tap1})  the covariance matrix   $\Sigma(\theta)$ is tapered, instead in  (\ref{eq:tap2})   the
$\Sigma(\theta)$ as well as the empirical  covariance matrix  $Z{Z}\,^\intercal$ are  tapered.
So the first approximation is computationally more efficient
nevertheless the derivative  of (\ref{eq:tap2}) leads to an unbiased estimating equation. For this reason
the  recent  literature
\citep{Shaby:Ruppert:2012, Stein:2013}
   has been focused on (\ref{eq:tap2}).


 \citet{Shaby:Ruppert:2012} show that, under increasing domain asymptotics \citep{Cressie:1993},  the maximizer of  (\ref{eq:tap2}) has asymptotic Gaussian distribution and
the asymptotic variance is given by the inverse of  the Godambe
information matrix
\begin{equation}\label{eq:god_tap}
G_{{T}}(\theta,d)=H_{{T}}(\theta,d)J_{{T}}(\theta,d)^{-1}H_{{T}}(\theta,d)^\intercal ,
\end{equation}
where
\begin{equation}
H_{{T}}(\theta,d)=-\E[\nabla^{2}l_{{T}}(\theta,d)],\qquad J_{{T}}(\theta,d)=\E[\nabla l_{{T}}(\theta,d)\nabla l_{{T}}(\theta,d)\,^\intercal ].
\end{equation}
The generic entries of the  $H_{{T}}(\theta,d)$  and $J_{{T}}(\theta,d)$ matrices are respectively:
$$[H_{{T}}(\theta,d)]_{ij}=\frac{1}{2}\mathrm{tr}\left\{B_i \left(\frac{\partial\Sigma(\theta)}{\partial \theta_j}
\circ R(d)\right)\right\}
$$
$$
[J_{{T}}(\theta,d)]_{ij}=\frac{1}{2}\mathrm{tr}
\left\{
\left[
B_i
\circ R(d)
\right]\Sigma(\theta)
\left[
B_j
\circ R(d)
\right]\Sigma(\theta)
\right\}
$$
where
$$B_i=[\Sigma_T(\theta)]^{-1}\left(\frac{\partial\Sigma(\theta)}{\partial \theta_i}
\circ R(d)\right)[\Sigma_T(\theta)]^{-1}.$$

Note that   $\lim_{d\rightarrow \infty}l^2_{{T}}(\theta,d)=l(\theta)$, that is when increasing the taper range  an improvement of the statistical efficiency is expected.  At the limit  the  asymptotic variance is given by the Fisher information matrix  \citep{Mardia:Marshall:1984})   whose generic entries are:
\begin{equation}\label{eq:fisher}
[I_{ML}(\theta)]_{ij}=\frac {1}{2}\textrm{tr}\left([\Sigma(\theta)]
^{-1} \frac{d\Sigma}{d\theta_i} [\Sigma(\theta)]
^{-1} \frac{d\Sigma}{d\theta_j}\right).
\end{equation}

\section{Composite likelihood estimation based on pairs}\label{sec:CL}

Let  $A_k$ be a marginal or conditional set of  the data, the composite likelihood (CL) \citep{Lindsay:1988} is an objective function defined as
a product of $K$ sub-likelihoods
\begin{equation}\label{eq:CL}
CL(\theta)=\prod_ {k=1}^K L(\theta; A_k)^{w_k},
\end{equation}
where  $L(\theta; A_k)$ is the likelihood generated from $f(z;\theta)$
by considering only the random variables in $A_k$
and $w_k$ are suitable non negative weights that do not depend on $\theta$.
The maximum CL estimate is given by $\hat\theta=\textrm{argmax}_{\theta}\, CL(\theta)$.

The choice of which and how many factors in (\ref{eq:CL}) can be  related to the  computational and
statistical
efficiency \citep{Lindsay:Yi:Sun:2011}.
Setting $A_k=(Z(s_i),Z(s_j))$, we obtain the  pairwise marginal Gaussian likelihood $L_{ij}$. If we let $A_k=(Z(s_i)|Z(s_j))$ we obtain
the pairwise conditional Gaussian likelihood  $L_{i|j}$ and finally setting  $A_k=(Z(s_i)-Z(s_j))$ we obtain the pairwise difference Gaussian likelihood $L_{i-j}$.
The computational cost for considering all possible pairs  is of  order $O(n^2)$ while it is of order $O(n^3)$ in
considering all possible  three-wise  $i.e.$ the same order of the evaluation of the likelihood for Gaussian  random fields. Thus from a computational point of view only the pairwise CL is opportune.

The expression for the logarithm of the sub-likelihoods are
\begin{eqnarray}
 l_{ij}(\theta)&=&-\frac{1}{2}\left\{2\log \sigma^2+\log(1-\rho^2_{ij})+\frac {B_{ij}} {\sigma^2(1-\rho^2_{ij} )}\right\}\label{eq:sub-lik-m}\\
l_{i|j}(\theta)&=& -\frac{1}{2}\left\{  \log \sigma^2+\log (1-\rho^2_{ij})+\frac {G_{ij}^2}{\sigma^2(1-\rho^2_{ij} )}\right\} \label{eq:sub-lik-c} \\
l_{i-j}(\theta)&=& -\frac{1}{2}\left\{ \log\sigma^2+ \log(1-
 \rho_{ij}) +\frac{U_{ij}^2}{2\sigma^2(1-
 \rho_{ij} )}\right\}\label{eq:sub-lik-d}
\end{eqnarray}
where $
 \rho_{ij}=\rho(s_i-s_j;\phi)$,
$B_{ij} =Z(s_i)^2+Z(s_j)^2-2 \rho_{ij} Z(s_i)Z(s_j)$, $G_{ij} = Z(s_i)- 2  \rho_{ij} Z(s_j)$,
$U_{ij}=Z(s_i)-Z(s_j)$.
The corresponding weighted composite log-likelihoods  are:
\begin{eqnarray}
pl_{M}(\theta)&=&\sum_{i=1}^n \sum_{j > i}^n l_{ij}(\theta) \label{eq:clp}w_{ij}
\label{eq:PCL},\\
 pl_{C}(\theta)&=&   \sum_{i=1}^n\sum_{j \neq i}^n l_{i|j}(\theta)w_{ij}
 =\sum_{i=1}^n \sum_{j > i}^n (2l_{ij}(\theta)-  l_{i}(\sigma^2)-  l_{j}(\sigma^2))w_{ij}
\label{eq:CCL},\\
 pl_{D}(\theta)&=&\sum_{i=1}^n \sum_{j > i}^n l_{i-j}(\theta)\label{eq:cld} w_{ij}
\label{eq:DCL},
\end{eqnarray}
where
$$l_{i}(\sigma^2)=-\frac{\log\,\sigma^2}{2}-\frac {Z(s_i)^2}{2\sigma^2} $$
 is  the  likelihood of one observation.
Here we  assume that the  weights are  symmetric, $i.e$ $w_{ij}=w_{ji}$.
Note that if the marginal parameter $\sigma^2$ is known, then marginal and the conditional pairwise likelihood have
the same efficiency. Otherwise it is not obvious which kind of estimation is more efficient.


A distinctive feature of $pl_a$, $a=C,D,M$,  is that the associated estimating function, $\nabla cl_a(\theta)$, is unbiased,
 irrespective of the distributional assumptions on the pairs.
In Appendix A we will show that the  maximum composite likelihood estimators  are  consistent and asymptotically normal.
Note that the above results have been derived in a more general settings than those in
\citet{Bevilacqua:Gaetan:Mateu:Porcu:2012} for  strictly
increasing sequence on evenly-spaced lattices. In contrast here  we do not impose any particular restrictions on the geometry and growth behavior of the lattice,
allowing unevenly spaced locations. This framework is more suited for real data analysis.

Under these results, again the inverse of the Godambe information matrix
\begin{equation}
G_{{a}}(\theta)=H_{{a}}(\theta)J_{{a}}(\theta)^{-1}H_{{a}}(\theta)^\intercal , \qquad a=C,D,M
\end{equation}
where
\begin{equation}
H_{{a}}(\theta)=-\E[\nabla^{2}pl_{{a}}(\theta)],\qquad J_{{a}}(\theta)=\E[\nabla pl_{{a}}(\theta)\nabla pl_{{a}}(\theta)\,^\intercal ]
\end{equation}
 is an approximation of the asymptotic variance of the CL estimator. In the Appendix B we can find closed form expressions  for the Godambe information.


The role of the weights in CL function is  to save computational time and improve the statistical efficiency.
In principle we can derive them using the theory on the optimal estimating functions \citep{Heyde:1997}.
Another possible strategy   is to consider different weight functions for different estimating equations coming from the CL  function. \citet{Bevilacqua:Gaetan:Mateu:Porcu:2012} showed that this kind of weights can improve the statistical efficiency of the method.
 A cut-off weight function  namely  $w_{ij}(d)=1$ if $\|s_i-s_j\|\le d$,
  and $0$ otherwise, has evident computational advantages with respect to a smooth one.
Moreover it can improve the efficiency  as it has been shown in  \citet{Joe:Lee:2009}, \citet{Davis:Yau:2011} and \citet{Bevilacqua:Gaetan:Mateu:Porcu:2012}.
The intuition behind this approach is that  the correlations between pairs of distant sampling locations are often nearly zero.   Therefore the use of the whole  pairs  may lose efficiency since too
many redundant pairs of observations can skew the information confined in pairs
of near observations. Hereafter we use $pl_a(\theta,d)$   to denote CL function based on pairs using simple cut-off weight function and $G_{{a}}(\theta,d)$ the associated Godambe information matrix.



\section{Numerical examples}\label{sec:numerical_examples}
\begin{figure}[h]
\begin{center}
\begin{tabular}{cc}
\includegraphics[width=0.5\textwidth]
{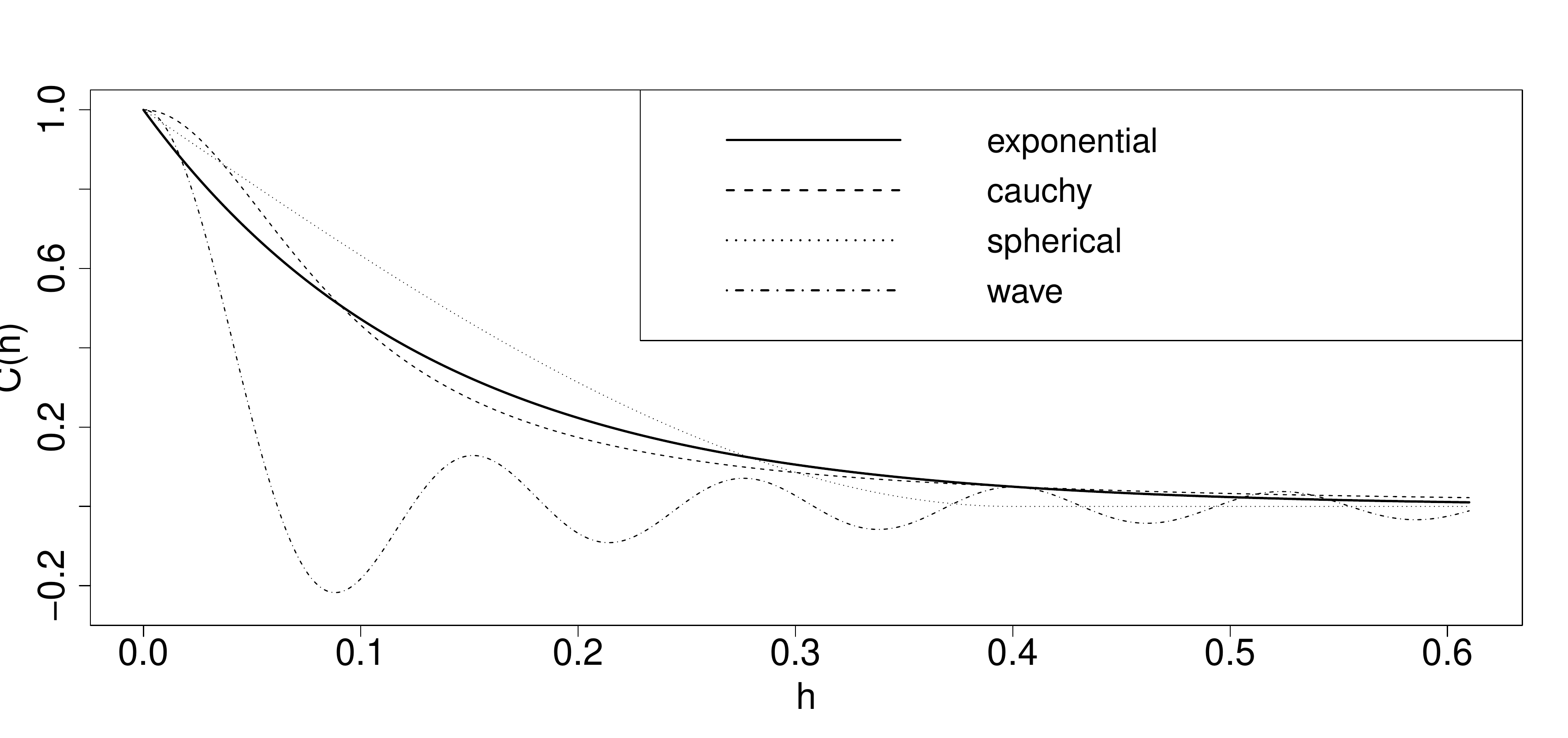}
&
\includegraphics[width=0.5\textwidth]
{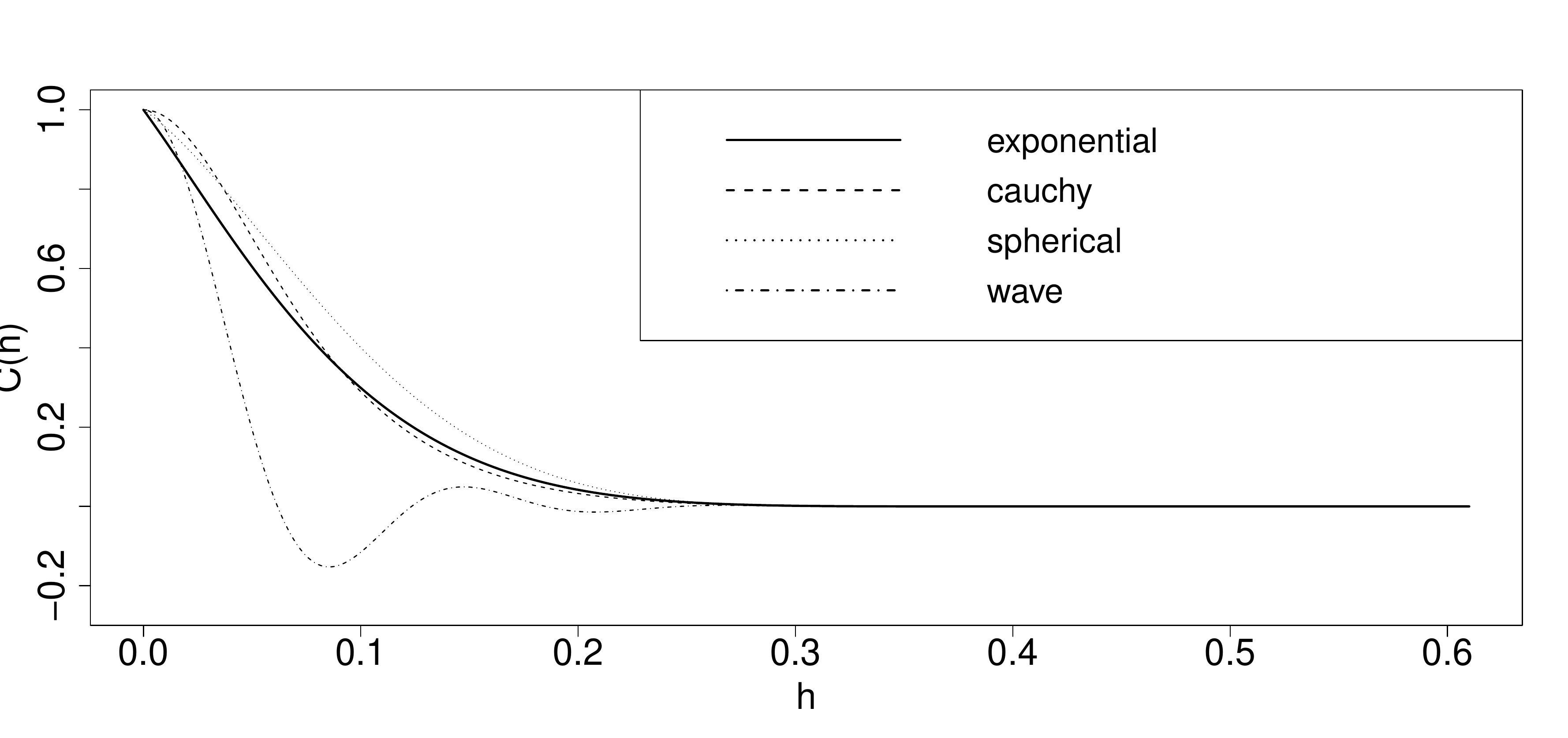}
\\
(a) & (b)
\end{tabular}
\end{center}
\caption{(a) Covariance functions with equivalent practical range, where $\sigma^2=1$, $\phi=0.4$, (b) the tapered covariance functions  using (\ref{eq:wendland}), with $d=\phi$}
 \label{fig:practical}
\end{figure}

We have considered four models:
\begin{enumerate}

\item
the exponential covariance function
\begin{equation}\label{eq:exp}
\displaystyle{
C(h;\theta)=\sigma^2\exp(-3 \|h\|/\phi)
}
\end{equation}
\item the Cauchy covariance  function
\begin{equation}\label{eq:cauchy}
\displaystyle{
C(h;\theta)=\frac{\sigma^2}{1+(\sqrt{19}\|h\|/\phi)^2}
}
\end{equation}
\item the spherical covariance function
\begin{equation}\label{eq:spherical}
 C(h;\theta) =
\left\{
\begin{array}{cl}
         \sigma^2(1 - 1.5  (\|h\|/\phi) + 0.5(\|h\|/\phi)^3) &  \textrm{if } \|h\| < \phi \\
         0  & \textrm{otherwise}
\end{array}
\right.
\end{equation}
\item the   wave or cardinal sine covariance function
\begin{equation}\label{eq:wav}
C(h;\theta) = \sigma^2(20.371\, \|h\|/\phi)^{-1}     \sin(20.371\, \|h\|/\phi)
\end{equation}
\end{enumerate}
The covariance models (\ref{eq:exp}),(\ref{eq:cauchy}) and (\ref{eq:wav}) are parametrized in terms of practical range that is
the correlation is lower than  $0.05$ when $\|h\|\geq  \phi$.
The aforementioned models cover a wide spectrum of situations that can arise in geostatistics.
The first model probably is the most commonly used model in geostatistics and it is a special case of the Mat\'ern
model  when $\nu=1/2$. Model (\ref{eq:cauchy}) is
 polynomially decreasing and hence more suitable than the exponential model for modeling of a slowly decaying covariance.
Model (\ref{eq:spherical}) is an example of compactly supported covariance  function, i.e. $C(h;\theta) =0$ for $\|h\| > \phi$. In principle
if  the taper range $d$ is greater than $\phi$ taper is not necessary. Model (\ref{eq:wav}) allows for negative correlations (see Figure \ref{fig:practical}).

For the  data locations we use the same setup as in
\citet{Kaufman:Schervish:Nychka:2008}.
We have considered   a regular grid   with increments $0.03$ over $W_k$
where $W_k=[0,2^{k/2}]\times [0,2^{k/2}]$,  $k=0,\ldots,5$. The grid points have been perturbed
adding a uniform random value on $[-0.01,0.01]$ to each coordinates  and, finally, we have randomly chosen $n_k=500 \cdot 2^k$ points without replacement.

First of all
 we compare the computational time required for one evaluation
of the likelihood (\ref{eq:loglik}), the tapered likelihood (\ref{eq:tap2}), the weighted  marginal pairwise likelihood (\ref{eq:clp}) and its unweighted version using the exponential covariance function, with $\phi=0.4$.
As taper  we consider the Wendland function (\ref{eq:wendland}).
As taper range and cut-off distance for the weighted composite likelihood estimator we have set $d=\phi$, i.e. the practical range . Figure \ref{fig:practical}  illustrates  the behaviour of the tapered correlation functions.



 Table \ref{tab:benchmark}
 depicts the saving in terms of computational burden for large datasets for
the marginal pairwise likelihood estimates (all calculations were carried out on a 2.27 GHz  processor with 6 GB of memory). In particular, the saving is quite remarkable when we consider the weighted version of the marginal pairwise likelihood.
Increasing $k$ and consequently the number $n$ of observations, the fraction of nonzero elements in the resulting tapered covariance matrix
decreases  (see the last column of Table \ref{tab:benchmark}).
Note also that for large taper range, i.e. small $n$, the overhead required for the sparse matrix overwhelms the expected computational advantages of the tapering estimator.
\begin{table}[th!]
\begin{center}
\begin{tabular}{rrrrrr}
  \hline
$n$ & ML &  $TAP(d)$ & $PL_M$ & $PL_M(d)$ & \% \\
  \hline
  500 & 0.38 & 0.34 & 0.03 & 0.00 & 0.33 \\
  1000 & 0.95 & 1.27 & 0.11 & 0.03 & 0.19 \\
  2000 & 3.21 & 4.77 & 0.40 & 0.05 & 0.11 \\
  4000 & 17.58 & 23.38 & 1.63 & 0.12 & 0.06 \\
  8000 & 129.83 & 120.02 & 6.53 & 0.29 & 0.03 \\
  16000 & 2388.16 & 789.69 & 26.82 & 0.78 & 0.01 \\
\end{tabular}
\end{center}
\caption{Time in seconds when evaluating ML, $TAP(d)$, $PL_M$ and $PL_M(d)$ functions  increasing the spatial domain of observation and the associated percentages of non zero values in the tapered matrix.}\label{tab:benchmark}
\end{table}

Now  we compare the asymptotic  relative efficiency  of the estimates   $\hat\sigma^2$ and $\hat\phi$  for the covariance models
 (\ref{eq:exp}), (\ref{eq:cauchy}), (\ref{eq:spherical}) and (\ref{eq:wav}).
As overall measure we consider:
\begin{equation}\label{eq:rel}
ARE_a(d)=
\left(
\frac{|G_a(\theta;d)|}{|I_{ML}(\theta)|}\right)^{1/p},\quad a=C,D,M,T
\end{equation}
where $I_{ML}(\theta)$ is the Fisher information matrix (\ref{eq:fisher}) and $p=2$ is the number of unknown components in $\theta$.
We have considered the case $k=0$ that is 500 locations sites over $[0,1]\times[0,1]$,  $\phi=0.4$ and  several increasing values of the taper range $d$, corresponding to
increasing values of the percentage of non zero values in the tapered covariance matrix.
Specifically we consider the sequence $0.01,0.02,\ldots, 0.2$ of percentages of non zero values.
The reason for considering this sequence is that, as outlined in \citet{Stein:Chen:Anitescu:2013},
 the tapered covariance matrix must be very
sparse to help a great deal with calculating the log determinant of the covariance matrix.

\begin{figure}
\begin{tabular}{cc}
\includegraphics[width=0.5\textwidth]
{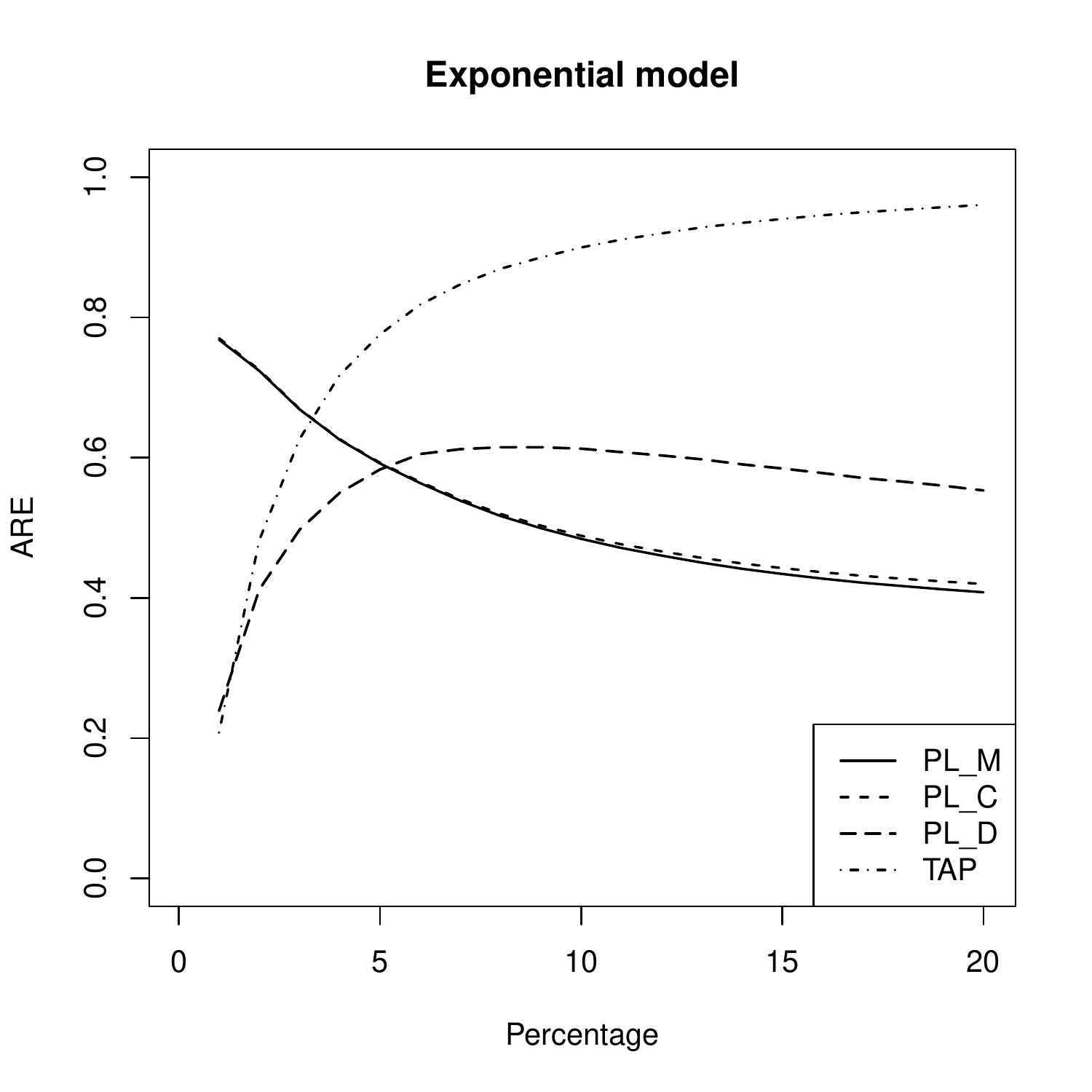}
&
\includegraphics[width=0.5\textwidth]
{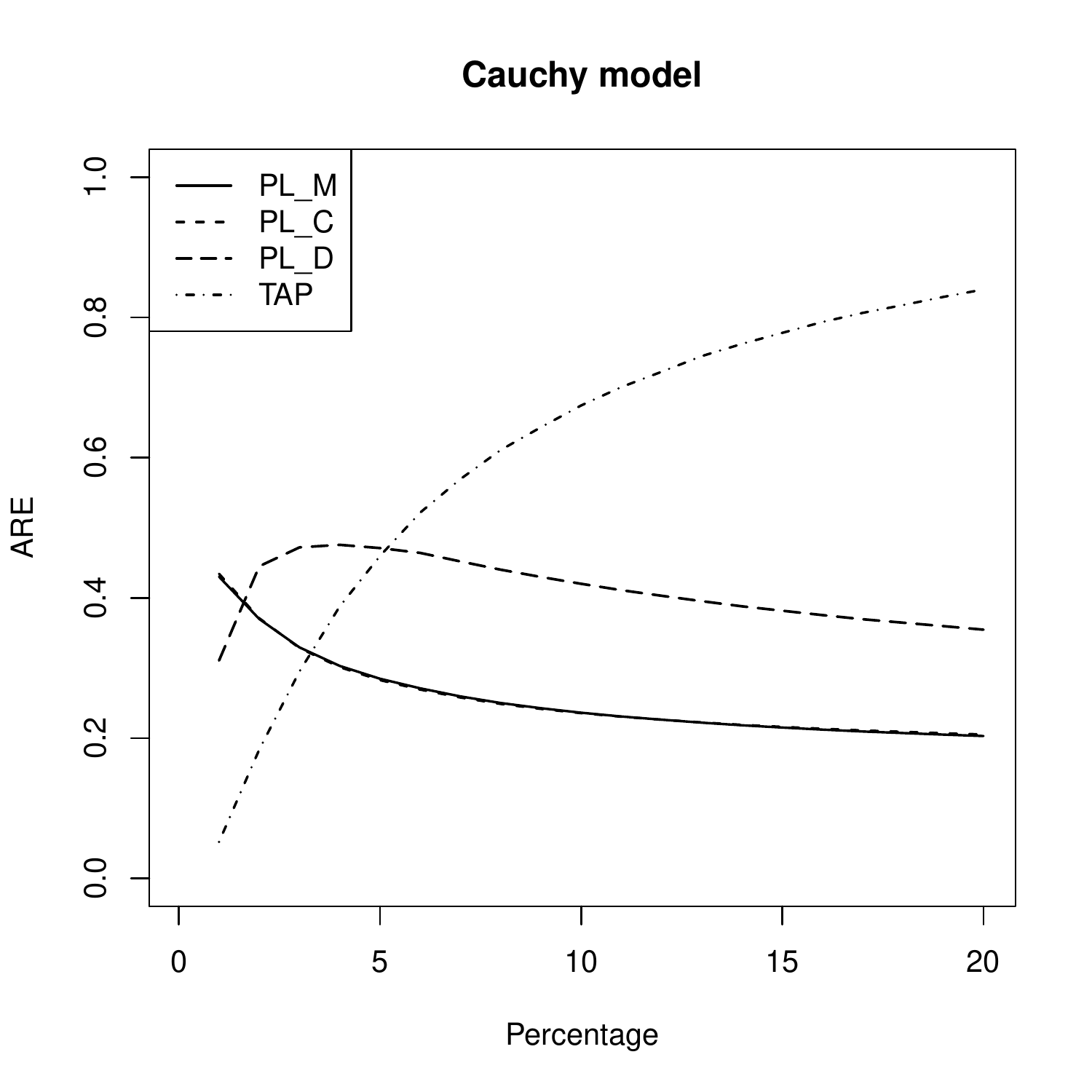}
\\
\includegraphics[width=0.5\textwidth]
{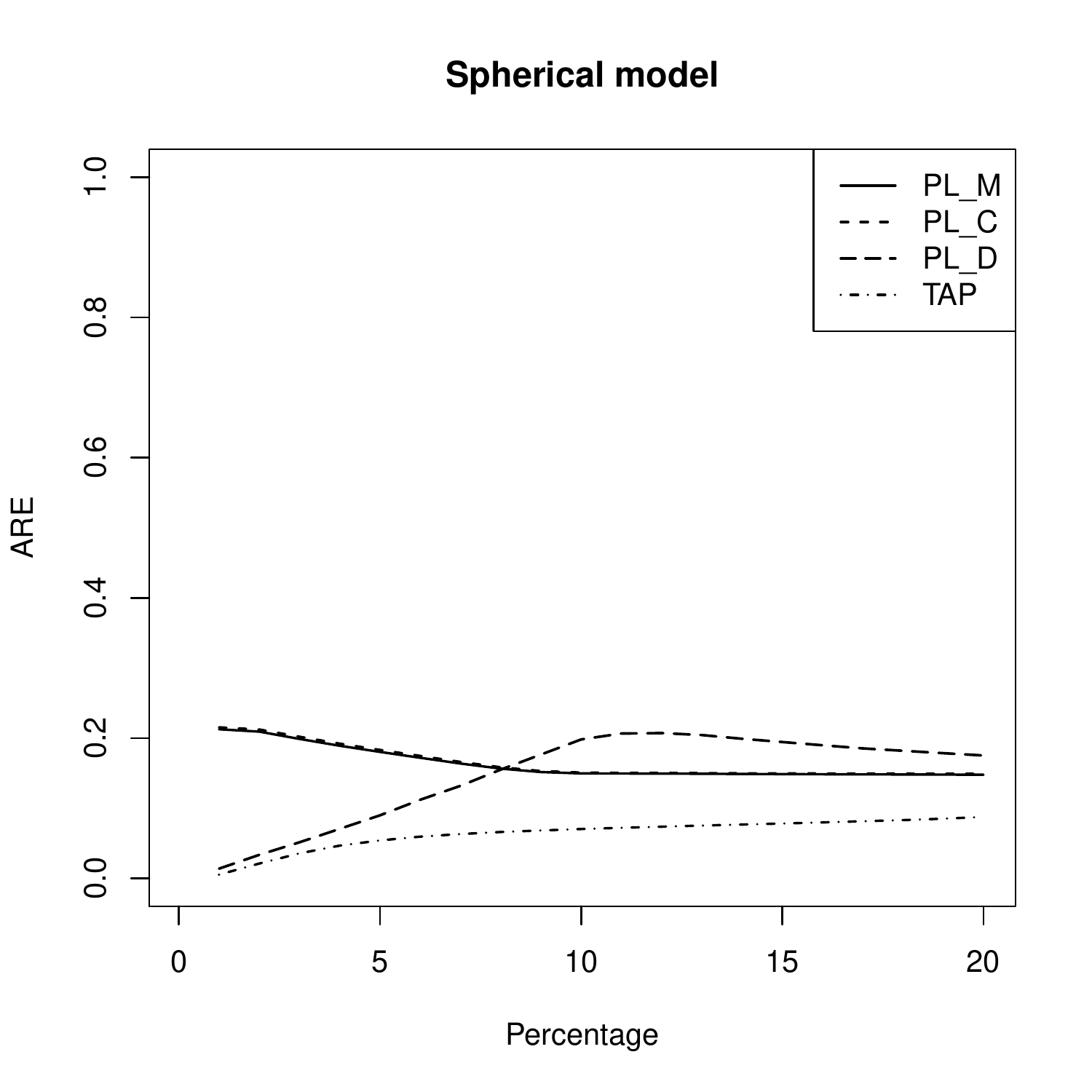}
&
\includegraphics[width=0.5\textwidth]
{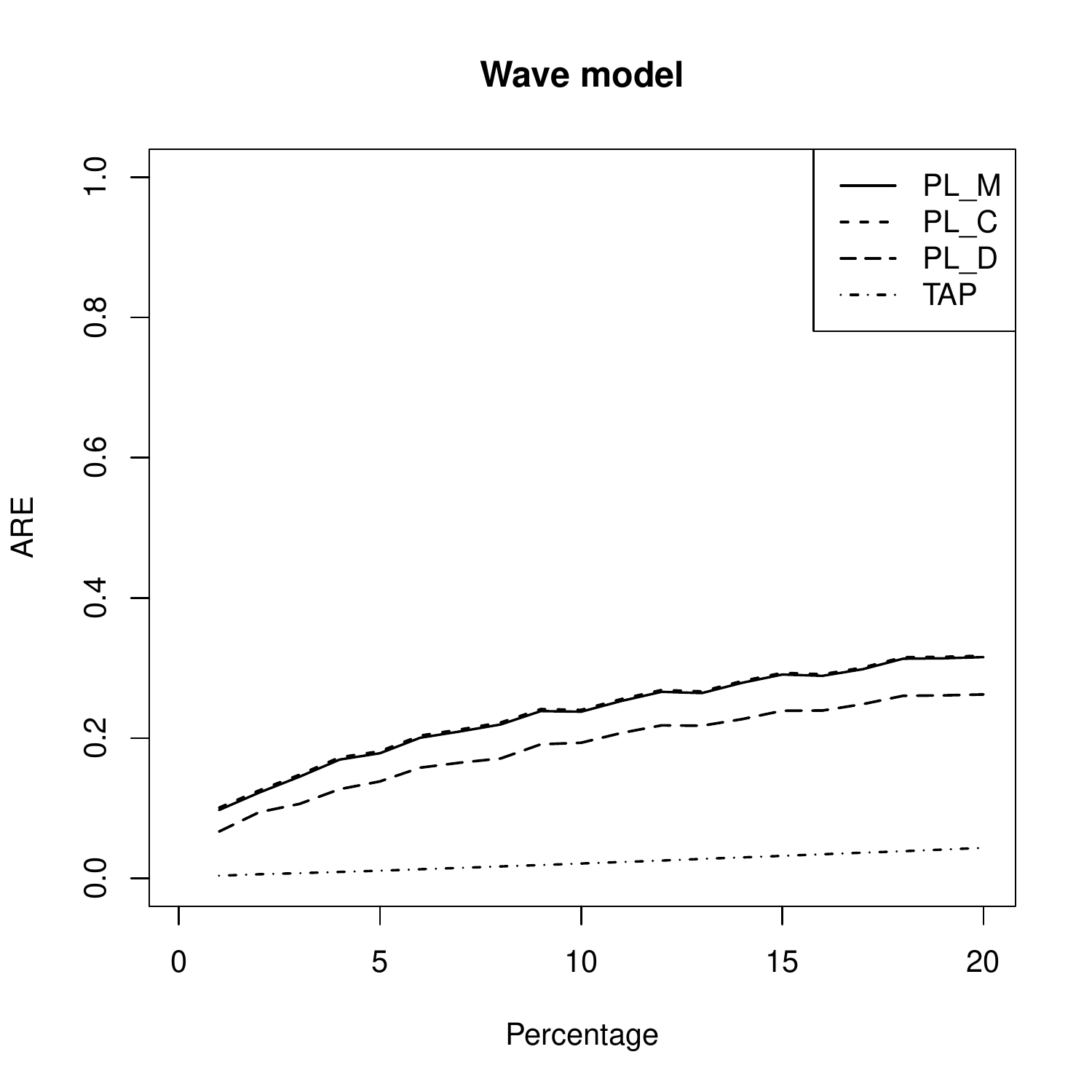}

\end{tabular}
\caption{ARE of the $TAP(d)$, and $PL_a(d),a=C,M,D$ estimators with respect of the percentage}
 \label{fig:are}
\end{figure}

In Figure \ref{fig:are} we depict  the measure  (\ref{eq:rel}) as a function
 of the considered percentages
of non zero values. As general remark for the taper method, the asymptotic relative efficiency is a monotonic increasing function of  the
percentages of non zero values as expected.

Moreover, for these examples, there is no practical difference in considering marginal and conditional  likelihood estimates, so
that a preference should be given to the first one because requires less computation.
On the other hand, for small percentages of non zero values, where the maximum tapered likelihood estimates takes advantage
from the sparsity of the covariance matrix,
 the maximum marginal and conditional pairwise likelihood estimates  outperform
the maximum tapered  likelihood  and the maximum difference pairwise likelihood estimates.

Note also that asymptotic efficiency of the maximum  CL  estimates is not a increasing function
of the distance considered in the weight function with the exception of the wave model.
These  examples suggest that a proper choice of the  distance $d$ can improve
significantly the statistical efficiency of maximum CL estimates under specific models.
Our findings add more evidence to previous results reported in the literature
\citep{Joe:Lee:2009,Davis:Yau:2011,Bevilacqua:Gaetan:Mateu:Porcu:2012}.
Moreover  such distance, i.e. the number of pairs,  in the  marginal and conditional pairwise CL should be different with respect to the distance of the difference  CL. For instance in the exponential model the `good' distance
for the marginal and conditional pairwise CL is approximately $0.06$ while for difference CL is $0.165$.

Looking at the behaviour for the different models, we see that the
the maximum tapered likelihood estimate performs reasonably well under the
exponential and the Cauchy model, but requires a large taper range for outperforming the maximum pairwise likelihood estimates,
vanishing the computational advantage of the sparsity of the covariance matrix.
Note that for the spherical and wave models the maximum tapered likelihood estimate perform worse than the CL methods bases on pairs
at least for the sequence of percentages of the tapered matrix considered.

Finally we  simulate 1000 random samples drawn from a Gaussian random fields under the same setting,
to  compare numerically the performance of CL methods based on pairs and tapering method with respect to ML one.
 All the estimates have been carried out using the R package
\texttt{CompRandFld} \citep{CRF:2012}, avalaible on  CRAN (\texttt{http://cran.r-project.org/}). In that package
all the estimation methods described here are fully implemented both in the spatial and spatio temporal case.
In particular the covariance tapering methods use the collection of R/Fortran functions  for sparse matrix algebra provided
by the package \texttt{spam} \citep{Furrer:Sain:2010}.

\begin{figure}
\begin{center}
\begin{tabular}{cc}
\includegraphics[width=0.50\textwidth]
{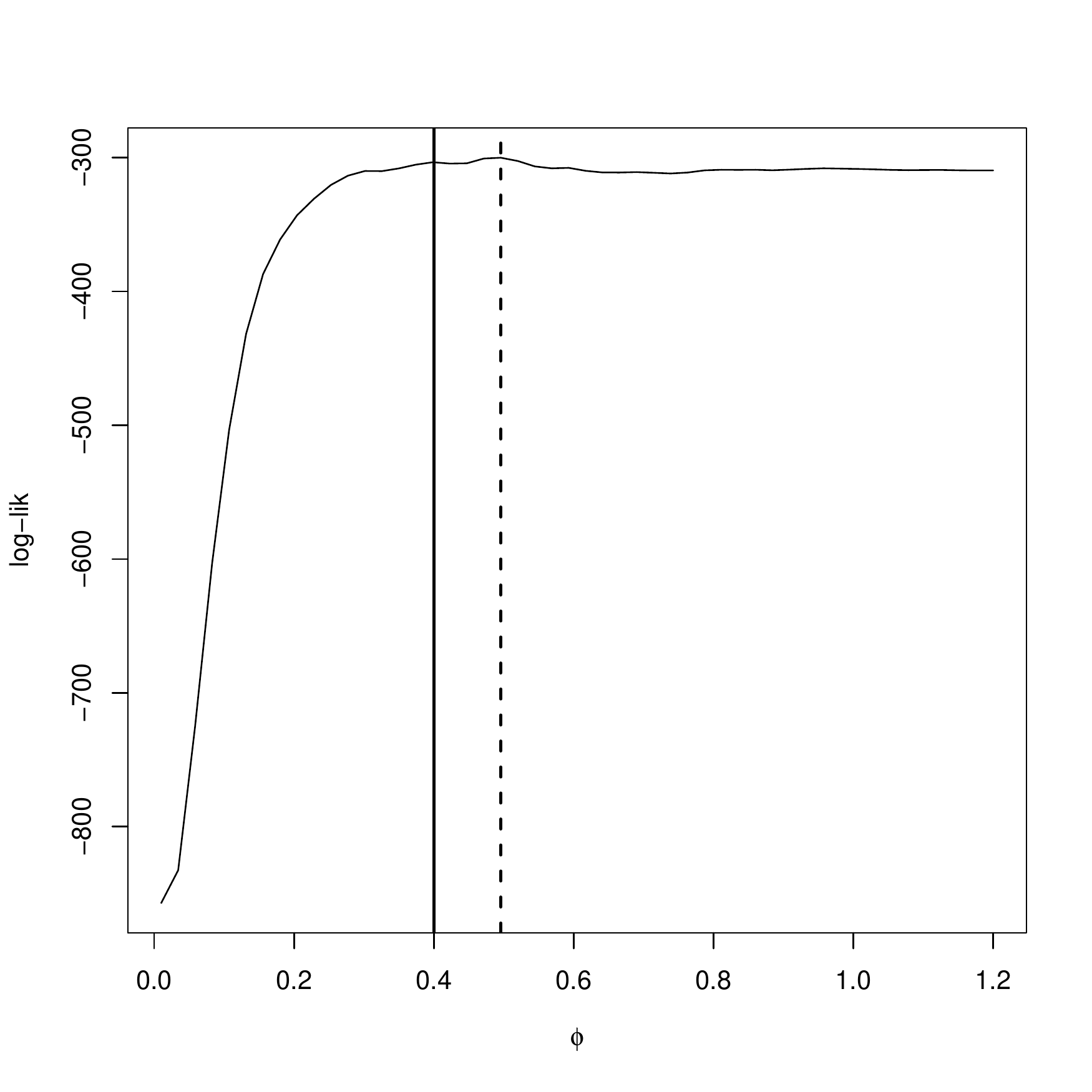}
&
\includegraphics[width=0.50\textwidth]
{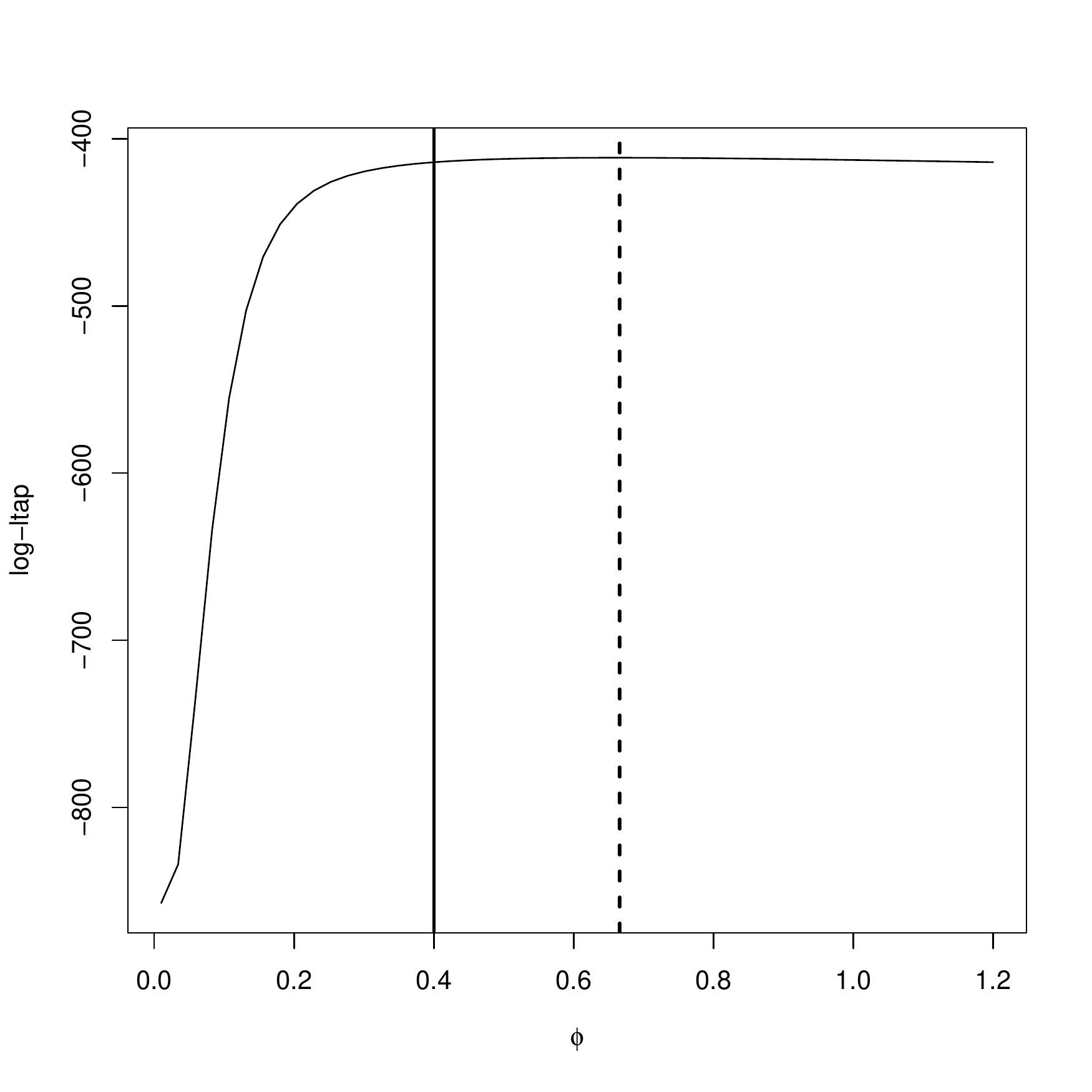}
\\
\includegraphics[width=0.50\textwidth]
{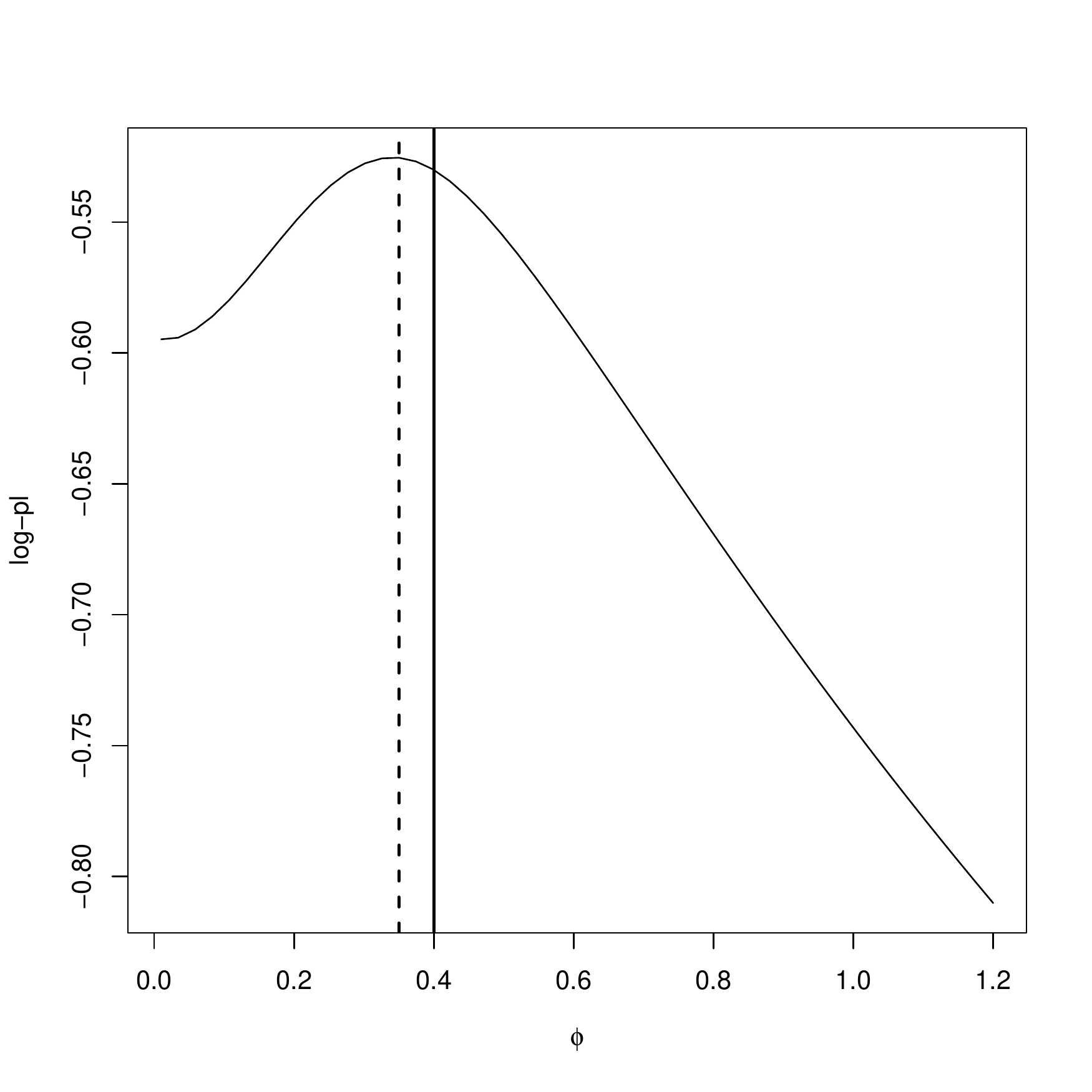}
&
\includegraphics[width=0.50\textwidth]
{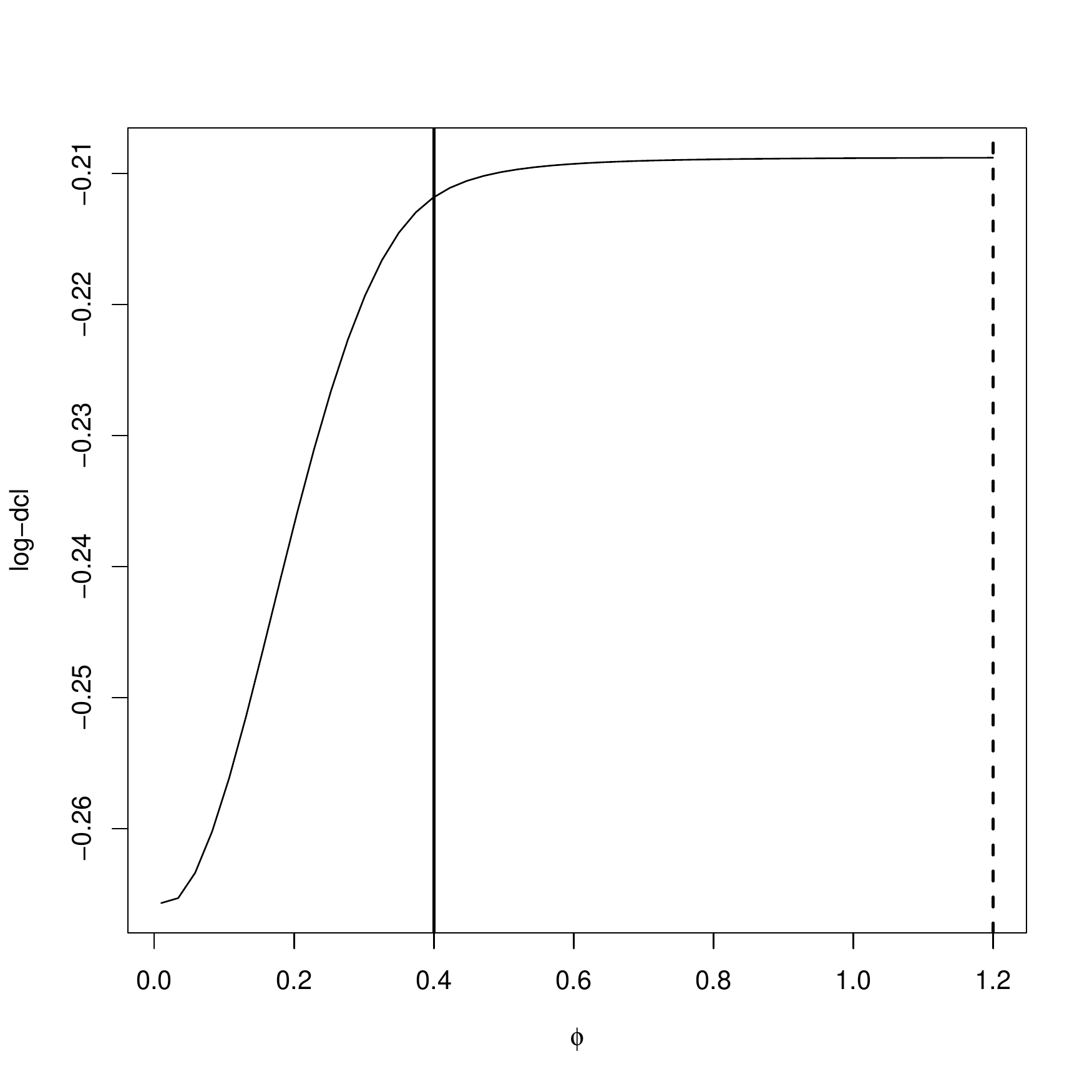}
\\
\end{tabular}
  \end{center}
\caption{Profile log functions with respect to  $\phi$ for the likelihood, tapering and CL based on pairs functions  for a particular simulation from a spherical model. In the plots
vertical solid lines correspond to the true value and  dashed lines are located to the maxima.
}
 \label{fig:profile}
\end{figure}

In the maximizing the objective functions for the spherical models, we have found some numerical difficulties.
In such case \citep{Mardia:Watkins:1989} reported that the log-likelihood may be multimodal for sample of finite size.
An example of these is given in Figure \ref{fig:profile}, where we plot the profile functions with respect to $\phi$, obtained in a simulation where the optimization procedure failed.

\begin{figure}
\begin{center}
\begin{tabular}{cc}
\includegraphics[width=0.50\textwidth]
{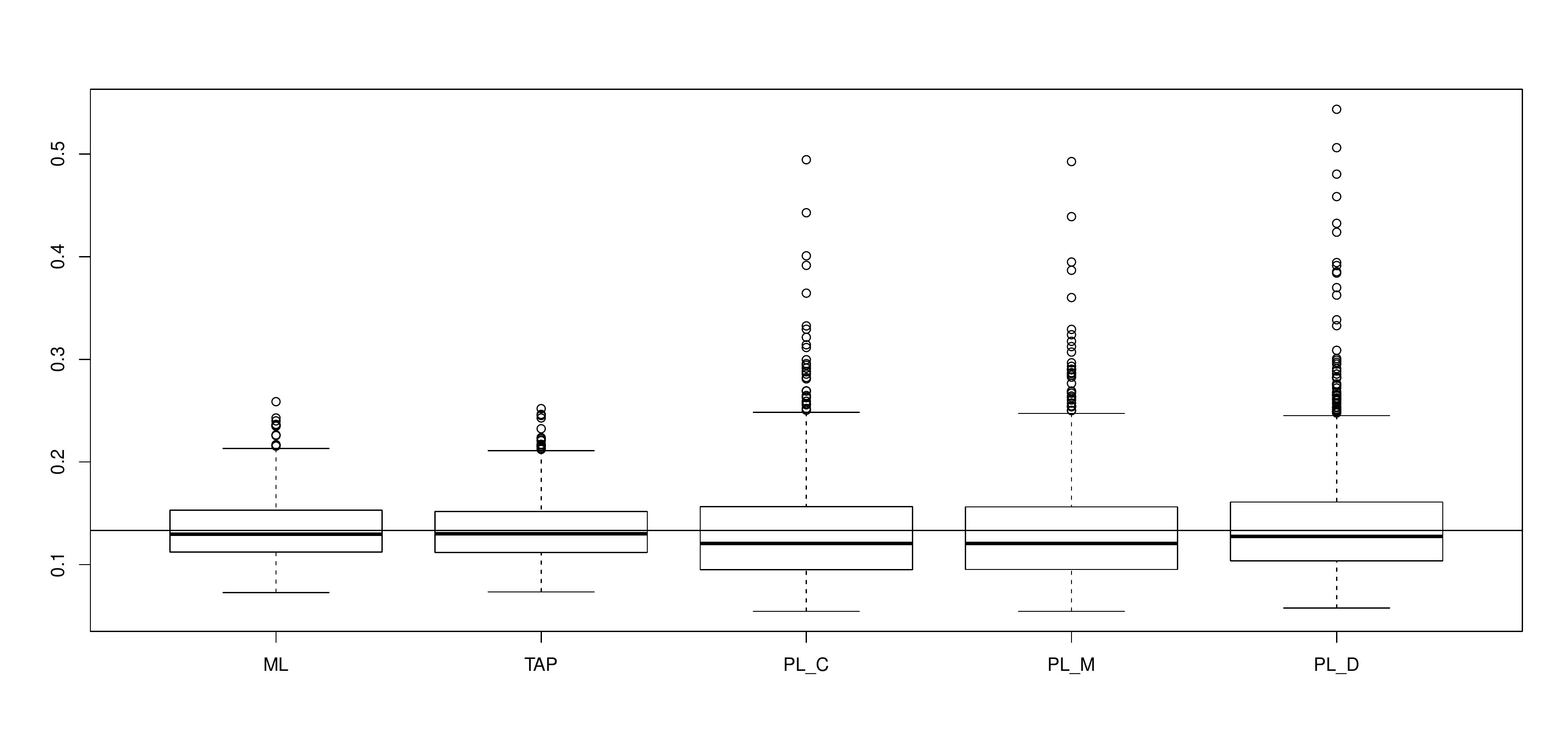}
&
\includegraphics[width=0.50\textwidth]
{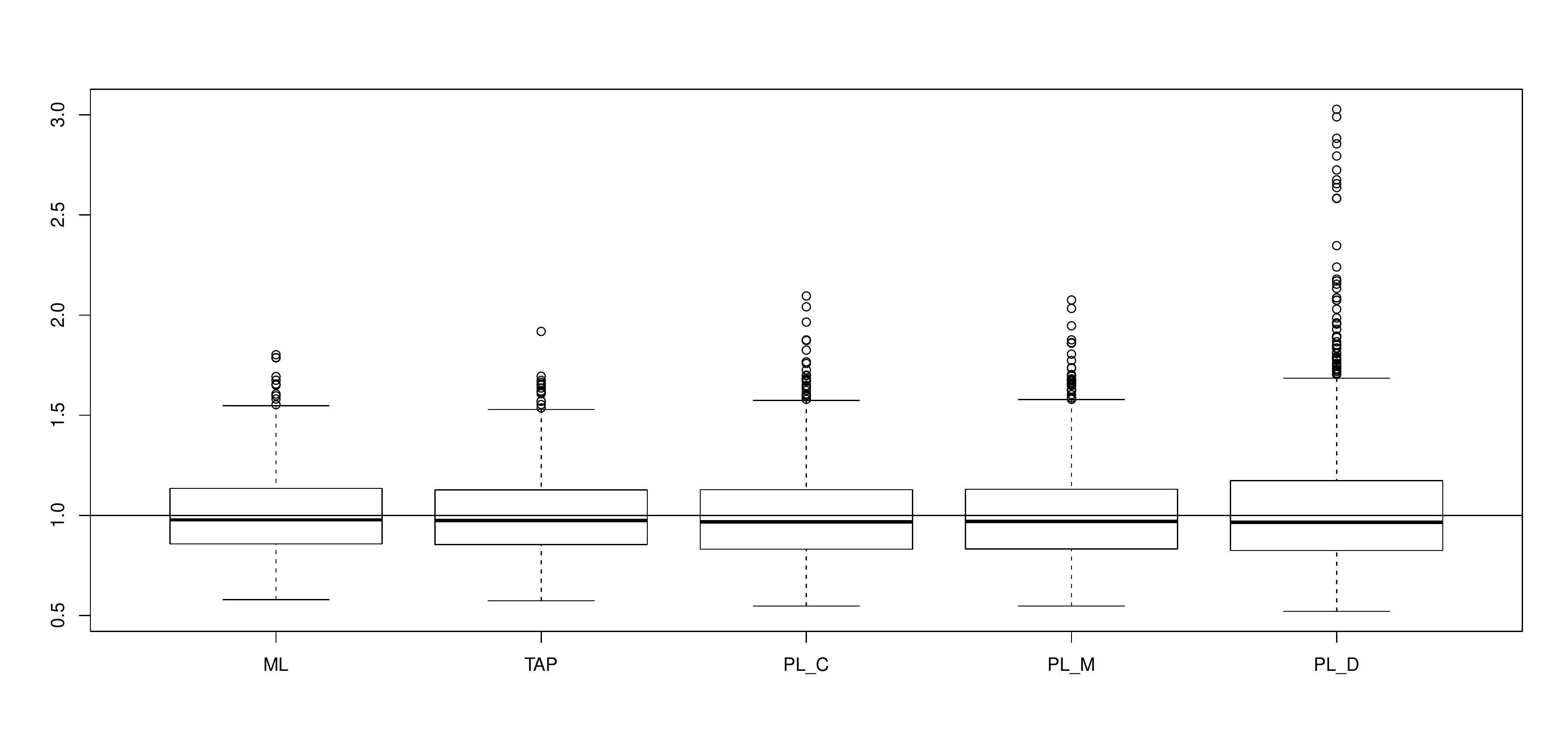}
\\
\includegraphics[width=0.50\textwidth]
{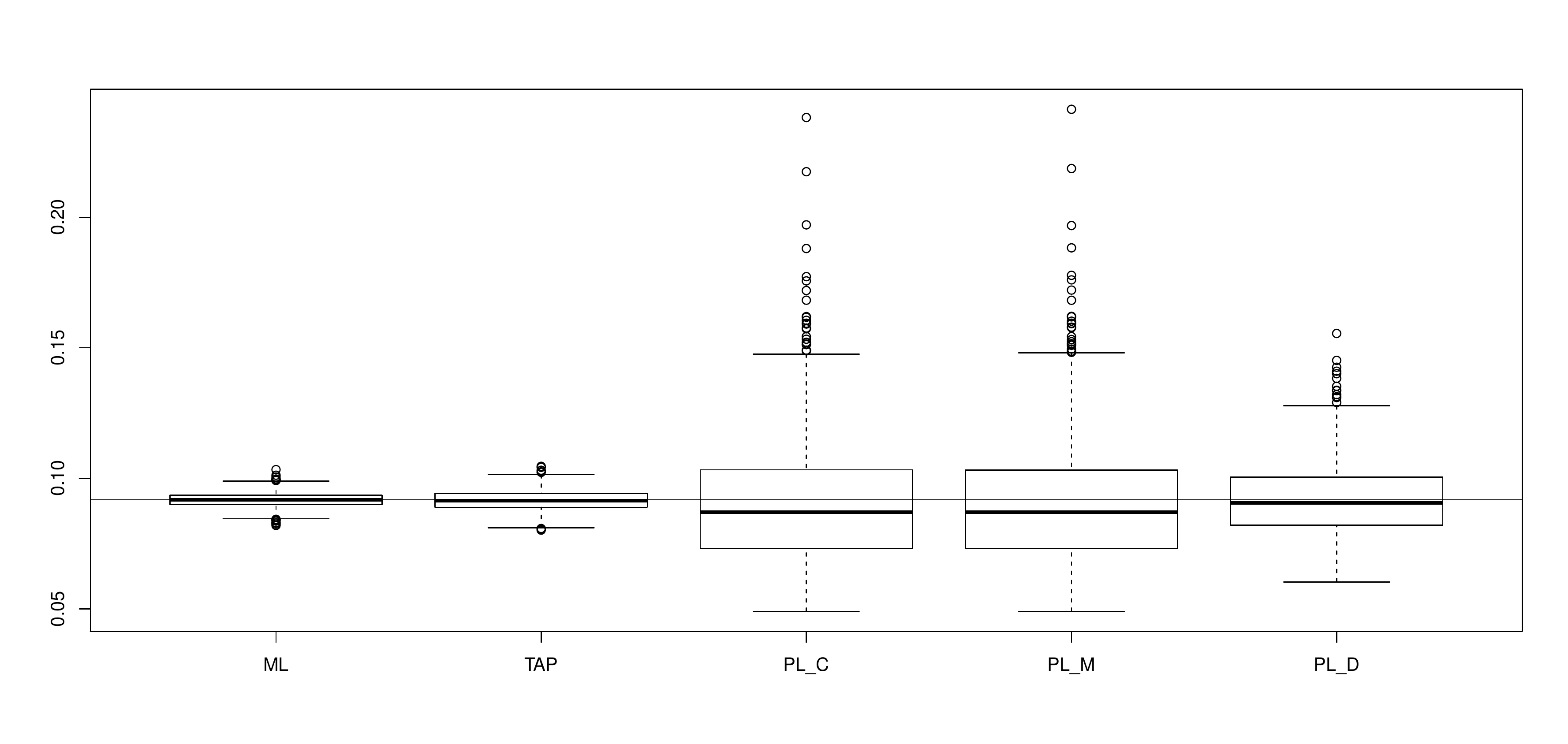}
&
\includegraphics[width=0.50\textwidth]
{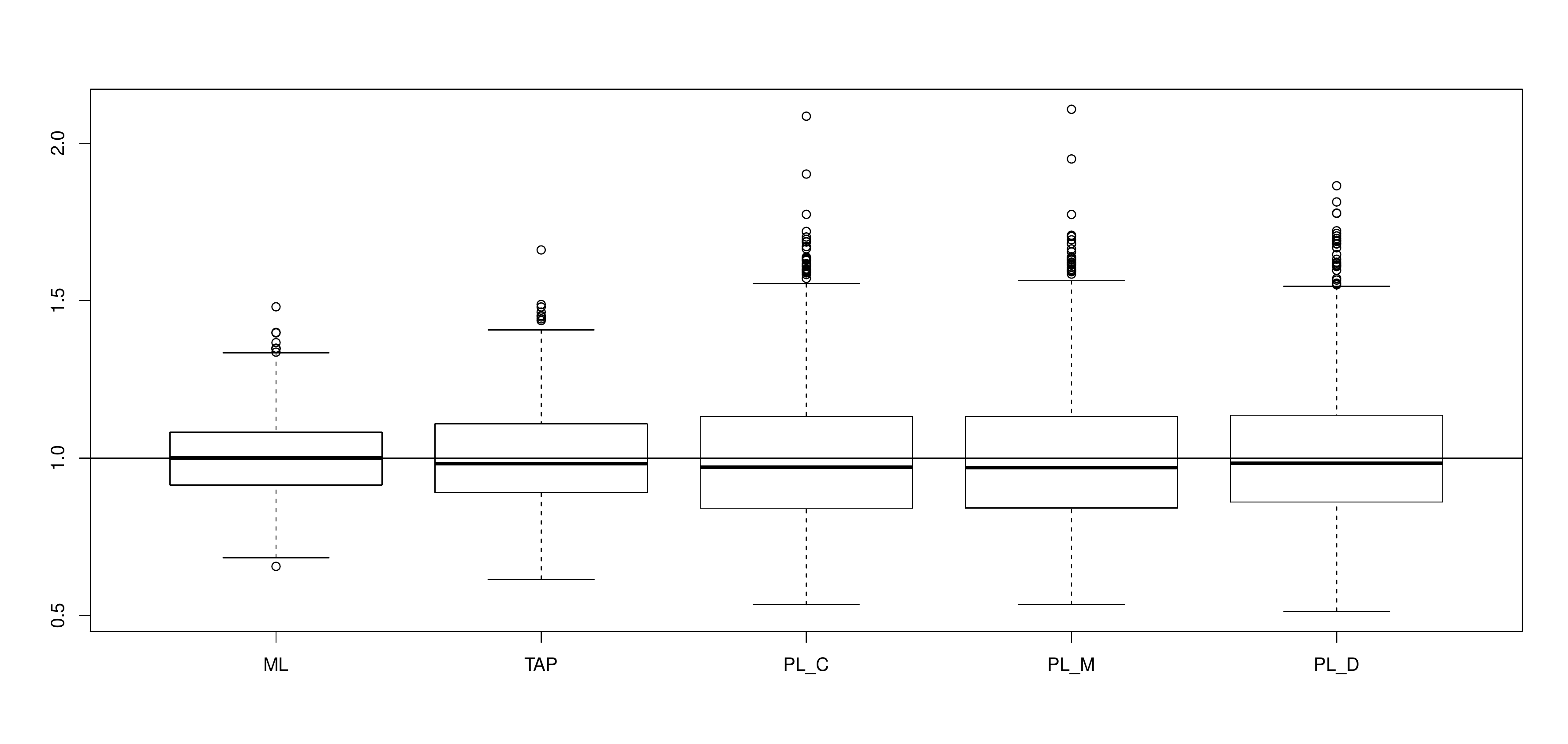}
\\
\includegraphics[width=0.50\textwidth]
{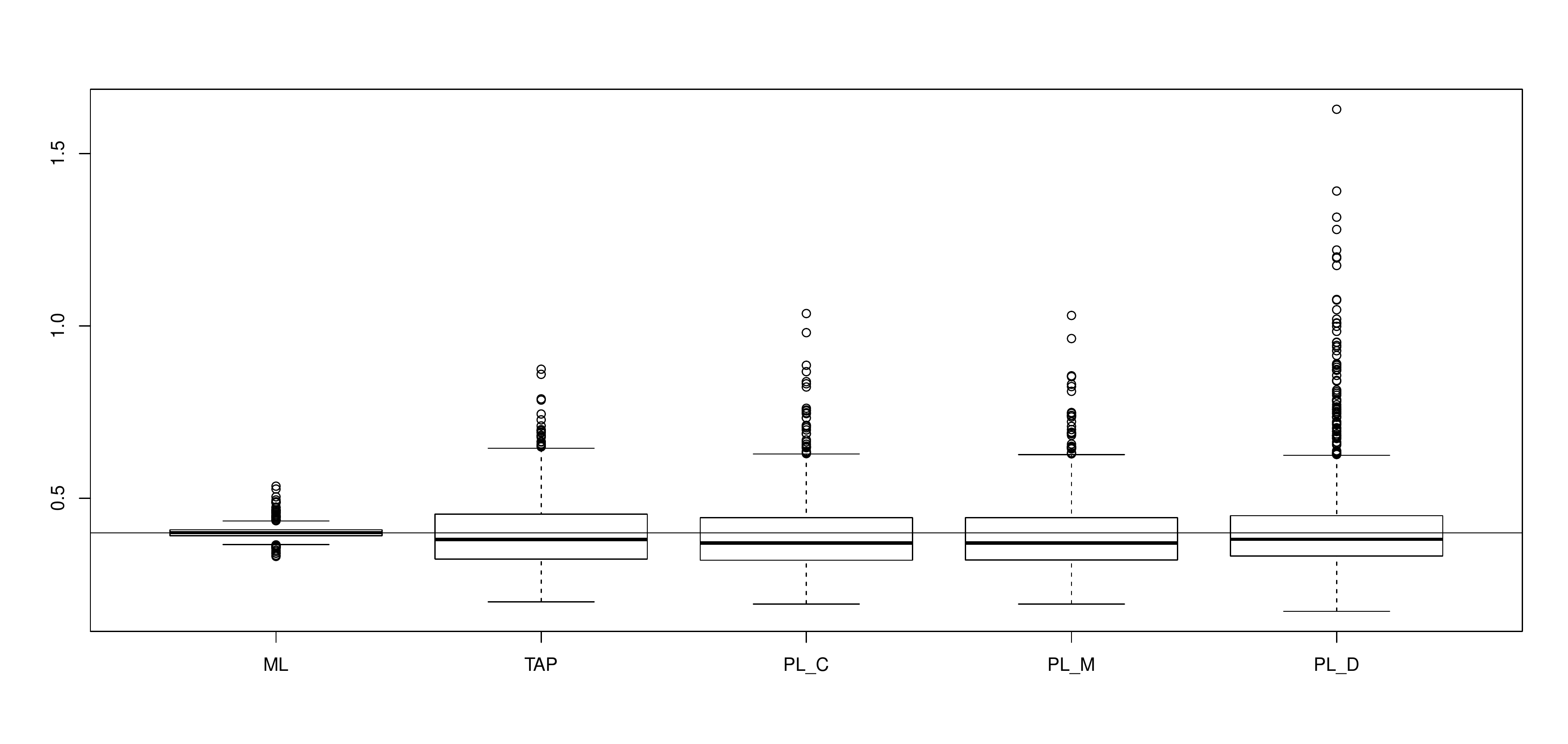}
&
\includegraphics[width=0.50\textwidth]
{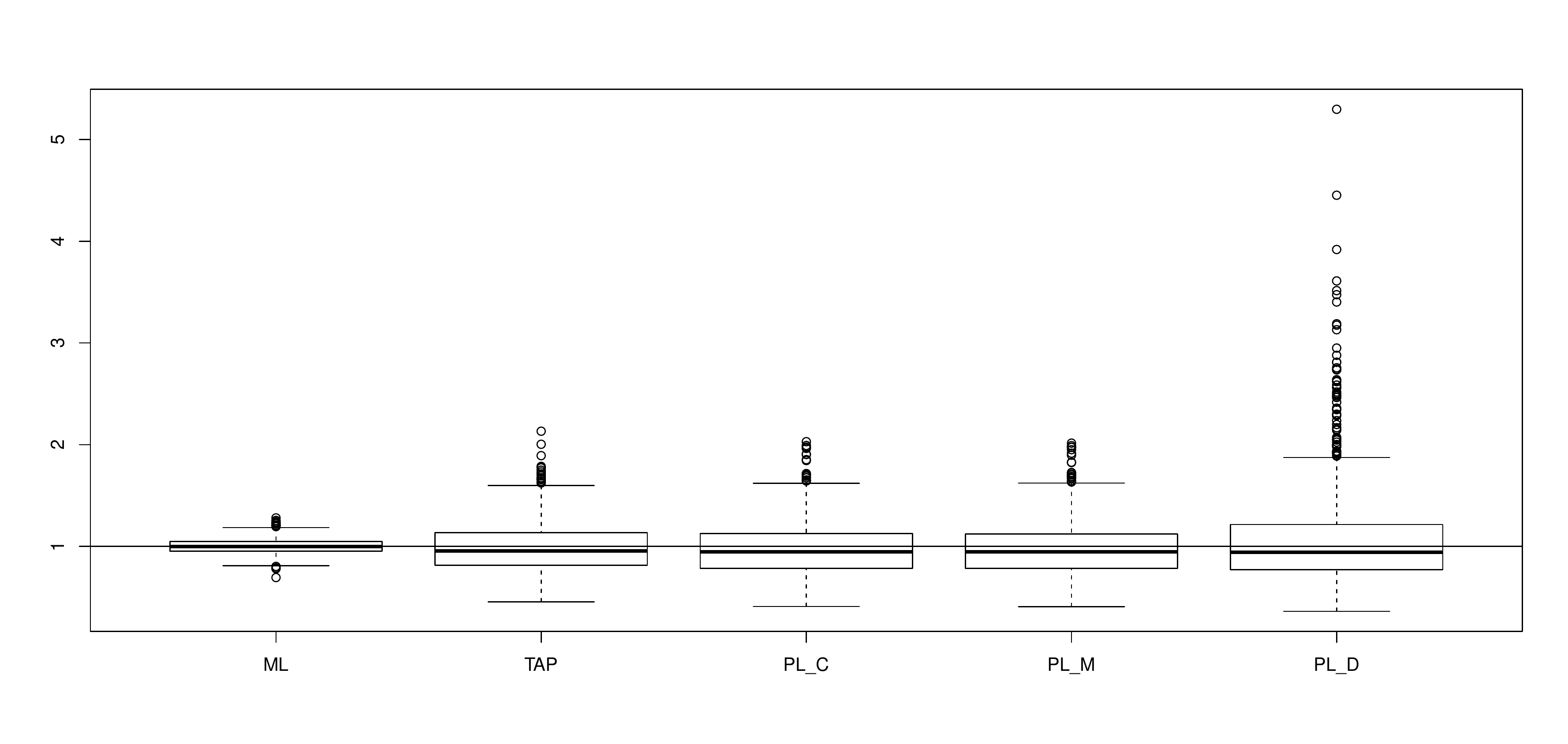}
\\
\includegraphics[width=0.50\textwidth]
{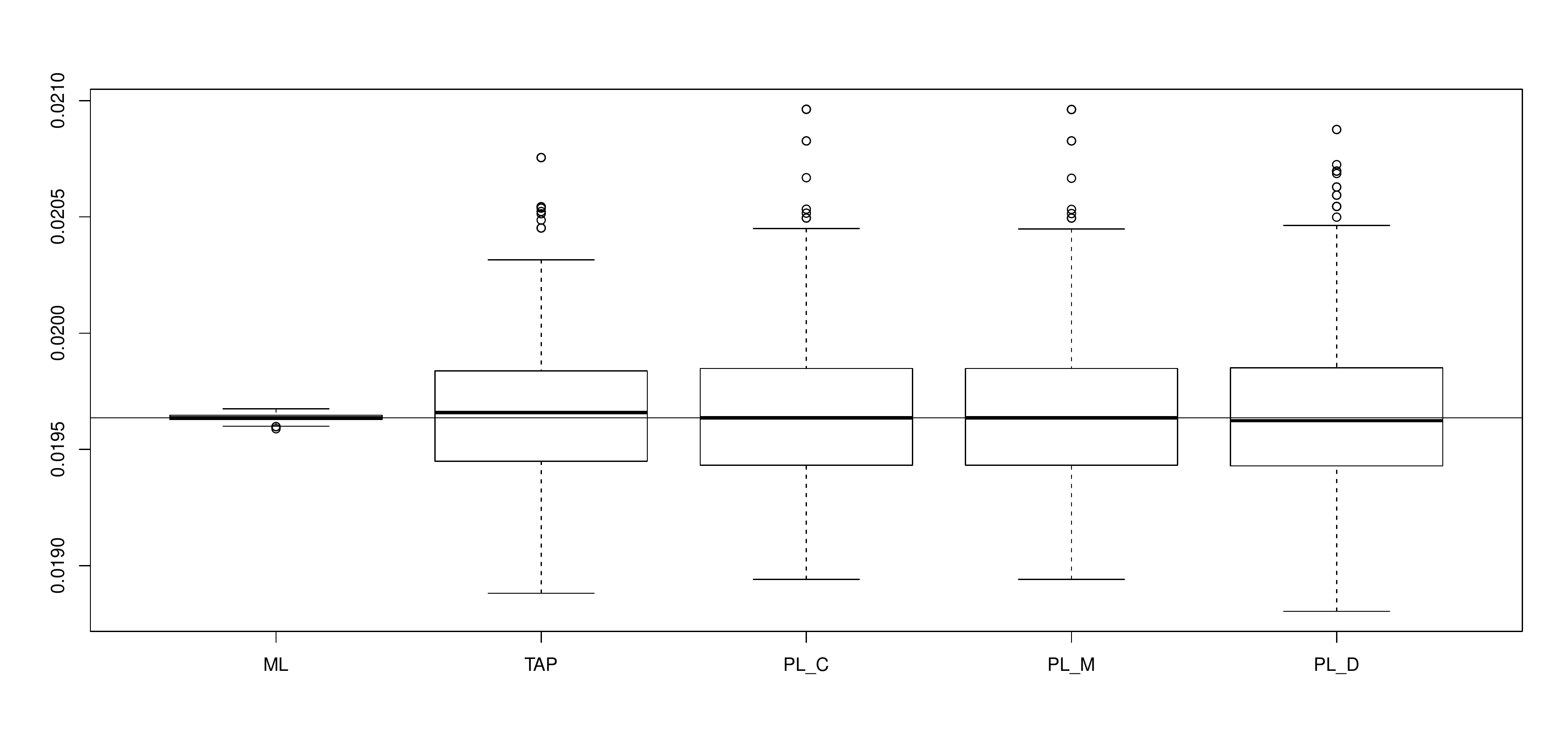}
&
\includegraphics[width=0.50\textwidth]
{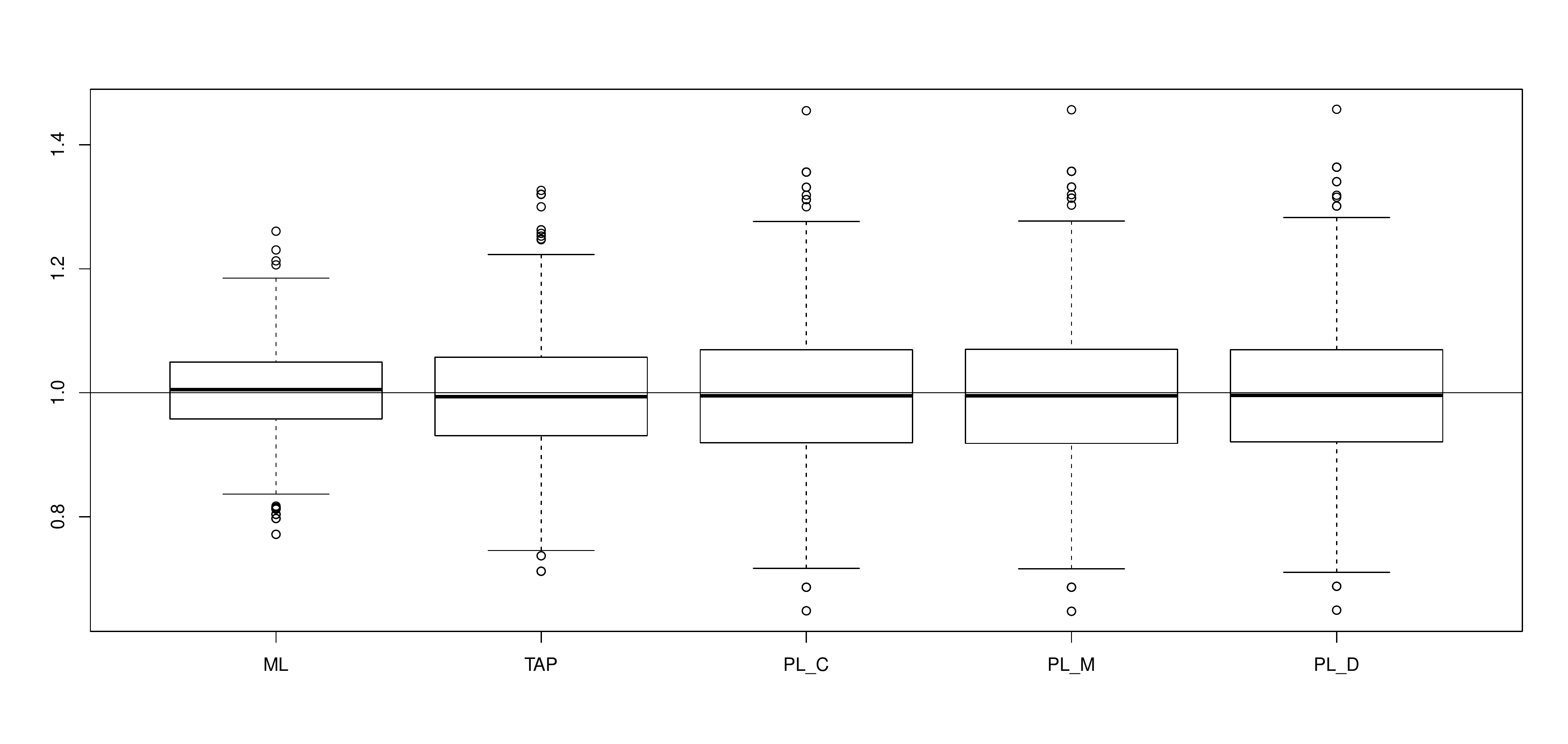}
\\
(a) & (b)
\end{tabular}
  \end{center}
\caption{Boxplots of the estimates [(a) $\phi$, (b) $\sigma^2$]. From the top, each row refers to the exponential, Cauchy, spherical and wave covariance model, respectively. In the boxplots the dashed lines represent the true values.
}
 \label{fig:boxplots}
\end{figure}

Note that the  marginal  pairwise likelihood is unimodal with a well-identified maximum value.
This is an example where, as outlined by  \citet{Varin:Reid:Firth:2011}, the composite likelihood surface can be much smoother than the full
joint likelihood, and thus easier to maximize.
The tapering  approach outperforms the CL methods (see Figure   \ref{fig:boxplots}) in the case of exponential and
Cauchy model, but shows the same performances for the other models.
Note also that the CL functions based on the differences yield to estimates with large variability.

\newpage
\begin{figure}[h!]
  \begin{center}
  \includegraphics[width=0.8\textwidth]{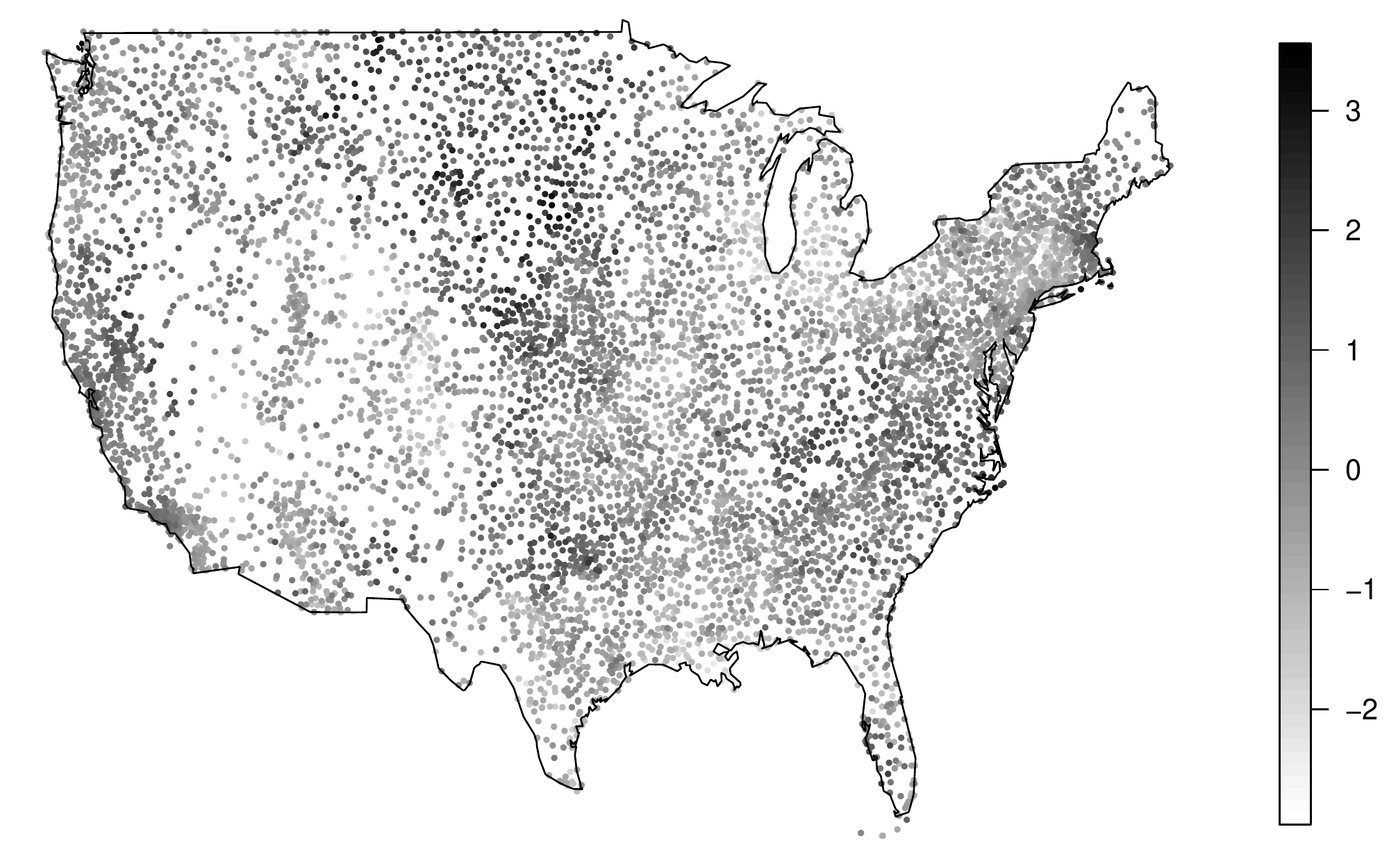}
  \end{center}
\caption{Yearly total precipitation anomalies registered  at  $7,352$ USA weather stations in 1962.} \label{fig:usa}
\end{figure}

\section{A real data example}\label{sec:dataset}
As data example we consider the data-set in \citet{Kaufman:Schervish:Nychka:2008} that
can be retrieved from
 \texttt{www.image.ucar.edu/Data/precip\_tapering/}. We consider
yearly total precipitation anomalies registered at  $7,352$ location sites in the USA  from 1895 to 1997 (see Figure \ref{fig:usa}).
The   yearly totals have been standardized by the
long-run mean and standard deviation for each station from 1962.
The data-set   can be considered of medium size allowing
 ML estimation  although it is very slow to compute.

\begin{figure}[h!]
  \begin{center}
\includegraphics[width=0.8\textwidth]{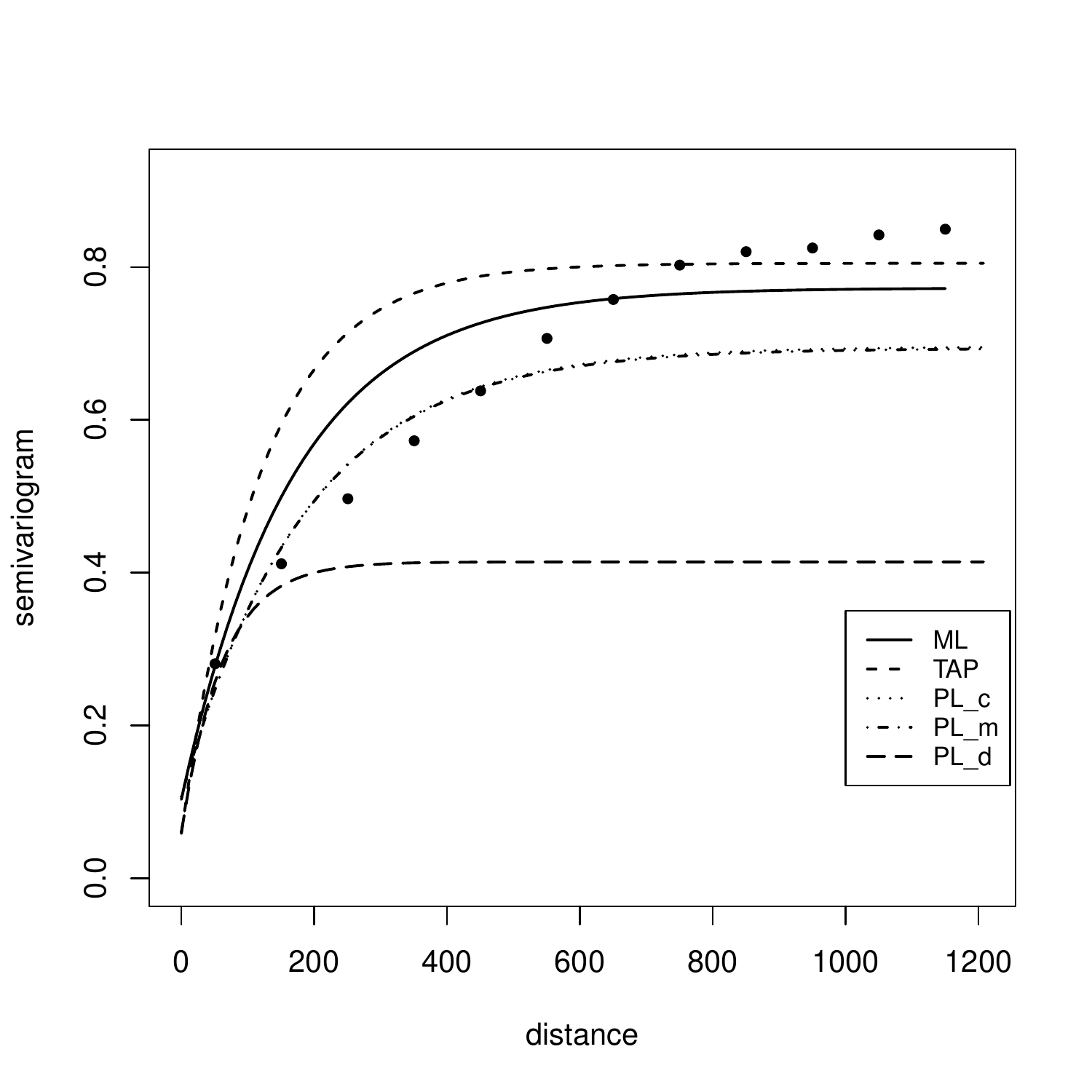}\\
  \end{center}
  \caption{Empirical variogram and theoretical ones,  estimated according to  different estimation methods.} \label{fig:vario_fit}
   \end{figure}

\citet{Kaufman:Schervish:Nychka:2008} adapted  a zero mean Gaussian random field with an exponential covariance model using the maximum likelihood  and the tapering method.
Here we choose an exponential covariance  model plus a nugget effect , i.e.
\begin{equation}\label{eq:model1}
C(h;\theta)=\tau^2I(||h||=0)+\sigma^2\exp\left\{-\frac{||h||}{\phi}\right\},
\end{equation}
as suggested by inspecting the empirical variogram   (Figure \ref{fig:vario_fit}).

The parameter $\theta=(\tau^2,\sigma^2,\phi)'$ is estimated  with  maximum likelihood, tapering and $pl_a(\theta;d)$, $a=C,D,M$  methods.
The distance between two sites are measured using the  great-circle distance  and the exponential function is still positive definite for this distance \citep{Huang:Zhang:Robenson:2011}. As taper function we use the  Wendland function (\ref{eq:wendland}) where the  taper range is fixed to $d=112.654$ Km,    the same value as in \citet{Kaufman:Schervish:Nychka:2008}.
Moreover we use  the same distance in $pl_a(\theta;d)$, $a=C,D,M$.

Table   \ref{tab:results}  reports the estimates  of maximum likelihood, tapering and $pl_a(\theta;d)$, $a=C,D,M$ methods and the associate standard errors. For maximum likelihood and tapering methods  standard errors   are computed  using the  inverse of the Fisher and Godambe information matrices in (\ref{eq:fisher}) and  (\ref{eq:god_tap}).
The evaluations of the plug-in estimates for $H_a$  and $J_a$ for the CL methods are of order $O(n^2)$ and $O(n^4)$ respectively, and the evaluation of $J_a$  become computationally unfeasible for large data-sets. For this  we  exploit the spatial subsambling techniques as explained  in \citet{Bevilacqua:Gaetan:Mateu:Porcu:2012}.
The results confirm the superiority of the tapering techniques in terms of efficiency, however $pl_a(\theta;d)$, $a=C,M$ are good competitors in terms of fitting (see Figure \ref{fig:vario_fit}).

\begin{table}[!h]
\begin{center}
\begin{tabular}[c]{|c|c|c|c|c|c|}
\hline
 \multicolumn{1}{|c|}{ \ }&$ML$ & $TAP(d)$ & $PL_C(d)$ &$PL_M(d)$&$PL_D(d)$\\
\hline
$\tau^2$          &$0.1033    $   &$0.0586$&    $ 0.1069  $        &      $ 0.1070$        &         $  0.06019$\\
                  &$(0.0042)  $    &$     (0.0088)  $&                 $  (0.0026)$&             $ (0.0033)  $        &        (0.0083)   \\
\hline
$a$               &$168.1174  $  &  $119.0561$     &  $186.2457 $      &    $185.7594 $       &          $62.25251$\\
                  &$ (12.2329)$  &   $(9.4117)$   &  $(17.3979)$    &     $ (19.0835) $  &          $  (13.7225)$   \\
\hline
${\sigma}^{2}$    &$ 0.6693   $  &  $0.7464$     &                           $  0.5890  $      &     $  0.5866 $        &        $   0.35391$\\
                  &$ (0.0632) $  &  $(0.0447)$    &     $(0.0182)   $ &       $  (0.0359) $        &        $ (0.0295)$ \\
\hline
\end{tabular}
\end{center}
 \caption{ $ML$, $TAP(d)$ and $PL_a(d),a=C,M,D$  estimates for the exponential covariance model with nugget effect
 (estimated standard errors are reported between parentheses).}\label{tab:results}
\end{table}

Next, we compare the prediction performance.
Assuming that $Z$ is a Gaussian random field, let $f_i$ the conditional
density of $Z(s_i)$ given   all observations  except $Z(s_i)$, i.e.
$$f_i(z)= \sqrt{2 \pi v_{-i}^2}
\,\exp
\left\{
-\frac 12\frac{\left(z-\widehat{Z}_{-i}(s_i)\right)^2 }{v_{-i}^2}
\right\}
$$
where $\widehat{Z}_{-i}(s_i)$ and $v^2_{-i}$ are the conditional expectation and variance. In calculating these last quantities for large dataset, we make use of the formula in  \citet{Zhang:Wang:2010}.

We have considered three predictive scores \citep{Gneiting:Raftery:2007}, namely
\begin{enumerate}
  \item the root-mean-square error (RMSE)
$$
\textrm{RMSE}=\left[n^{-1}\sum_{i=1}^n\{Z(s_i)-\widehat{Z}_{-i}(s_i) \}^2\right]^{1/2}
$$
\item the logarithmic score (LSCORE)
$$
\textrm{LSCORE}=-n^{-1}\sum_{i=1}^n\log f_i(Z(s_i))
$$
\item  the continuous ranked probability score (CRPS)
$$
\textrm{CRPS}= n^{-1}\sum_{i=1}^n\int_{-\infty}^{\infty} [F_i(z)-I_{(\-\infty,z]}(Z(s_i)]^2dz
$$
\end{enumerate}
where $F_i (z)$ is the conditional cumulative distribution function of $Z(s_i)$

In Table   \ref{tab:pred1},  \ref{tab:pred2} and \ref{tab:pred3}   we report  RMSE, LSCORE and CRPS for the exponential model and we contrast them with an exponential model  without nugget effect, as proposed by
\citet{Kaufman:Schervish:Nychka:2008}.
Our findings highlight how we have an effective improvement when we consider an additional nugget effect.
Moreover $pl_a(\theta;d)$, $a=C,M$  estimates provides comparable results with respect to the
tapering method.
\begin{table}
\begin{center}
\begin{tabular}{|c|c|c|c|c|c|}
 \hline Model                       &$ML$        &$TAP(d)$&        $PL_C(d)$ &      $PL_M(d)$ &    $PL_D(d)$  \\
\hline Exponential with nugget      &0.2178264  &0.2208672  &  0.2177944     &0.2177954  &0.2196979 \\
\hline Exponential without nugget   &0.2293623 &0.2292307   & 0.2312318     &0.2312022   &0.231374 \\
\hline
\end{tabular}
\end{center}
\caption{Prediction performance in terms  of MSPE for exponential covariance model   with and without nugget effect estimated with $ML$, $TAP(d)$ and $PL_a(d),a=C,M,D$ methods.}\label{tab:pred1}
\end{table}

\begin{table}
\begin{center}
\begin{tabular}{|c|c|c|c|c|c|}
 \hline Model       &$ML$ &$TAP(d)$&$PL_C(d)$ &$PL_M(d)$ & $PL_D(d)$  \\
\hline Exponential with nugget    & 0.6380736 &0.6404574   & 0.6422388  & 0.6422617&0.6432923 \\
\hline Exponential without nugget &0.6773373  & 0.6697904  &  0.8676976 & 0.8686335  &0.696967 \\
\hline
\end{tabular}
\end{center}
\caption{Prediction performance in terms  of LSCORE for exponential covariance model   with and without nugget effect estimated with $ML$, $TAP(d)$ and $PL_a(d),a=C,M,D$ methods.}\label{tab:pred2}
\end{table}

\begin{table}
\begin{center}
\begin{tabular}{|c|c|c|c|c|c|}
 \hline Model                     &$ML$ &$TAP(d)$&$PL_C(d)$ &$PL_M(d)$ & $PL_D(d)$  \\
\hline Exponential with nugget    &0.446379&0.4470558  &  0.4438936   & 0.4438985 &0.4436438 \\
\hline Exponential without nugget &0.455952 & 0.4607079  &  0.4395014  & 0.4394311 &0.4529027\\
\hline
\end{tabular}
\end{center}
\caption{Prediction performance in terms  of CRPS for exponential covariance model   with and without nugget effect estimated with $ML$, $TAP(d)$ and $PL_a(d),a=C,M,D$ methods methods.}\label{tab:pred3}
\end{table}

\section{Concluding remarks}\label{sec:conclusions}

The class of CL  functions is very large and for a given estimation problem it is not clear how to choose in this class.
In the Gaussian case, if the choice of the CL is driven by computational concerns
then the CL based on pairs have clear computational advantages with respect to other type of CL.

In this paper through theoretical and numerical examples we have compared
three methods of weighted CL based on pairs (marginal, conditional and difference), using the
 the covariance tapering method as benchmark.

One advantage of the covariance tapering method is that it is possible to control the trade-off between the statistical and computationally
 efficiency with the taper range while this is not the case for CL based on pairs as explained in the theoretical examples.
 The choice of the distance in the weight function for the CL based on pairs should be driven by some  optimality criterion.
 \citet{Bevilacqua:Gaetan:Mateu:Porcu:2012} for instance proposed a method based on the minimization of the trace of the Godambe information matrix. Nevertheless this method could be computationally hard in particular for large data-set.

 The theoretical and numerical examples highlights a slightly better performance of the  weighted conditional and marginal CL with respect to weighted difference CL.
 Moreover the weighted marginal CL are computationally preferable with respect to the tapering method while the tapering method
 shows better statistical efficiency when increasing the taper range.
Our suggestion for the practitioners is to consider both the methods when it is computationally feasible, as in the real data example proposed.
For  data sets of large dimension CL based on pairs is preferable since in general a little loss of statistical efficiency
is offset by good computational performances.
 Our findings are consistent with those of  \citet{Stein:2013}  who compares the covariance tapering with a specific type of composite likelihood based on independent blocks.

\section*{Appendix A}
Asymptotic results can be been proved for spatial processes which are observed
at finitely many locations in the sampling region.
In this case we deal with an  increasing domain  setup where the sampling region is unbounded.

We consider a  weakly dependent random field
$\{Z(s), s\in S\}$,
 defined over an  arbitrary lattice $S$ in  $\mathbb{R}^{d}$ that is
not necessarily regular. The lattice $S$ is
equipped with the metric $\delta(s_k,s_l)=\max_{1\le l \le d}|s_{i,l}-s_{j,l}|$ and the distance between any subsets $A,B\subset S$ is defined as
$\delta(A,B)=\inf\{\delta(s_k,s_l): \, s_k\in A\textrm{ and } s_l\in B\}$.
We denote
\begin{eqnarray*}
 \alpha(U,V)=\sup_{A,B}\{|P(A\cap B)-P(A)P(B)|:\, A\in
\mathcal{F}(U), B\in
\mathcal{F}(V)\},
\end{eqnarray*}
where $\mathcal{F}(E)$ is the
$\sigma$-algebra generated by the random variables $\{Z(s),\,
s\in E\}$.
The $\alpha$-mixing coefficient \citep{Doukhan:1994}  for the random field $\{Z(s), s\in S\}$  is defined as
$$
\alpha(a,b,m)=\sup_{U,V}\{\alpha(U,V),\,|U|< a,\, |V| <b , \,\delta(U,V) \ge m\}.
$$

We make the following assumptions:

\renewcommand{\labelenumi}{C\arabic{enumi}:}
\renewcommand{\labelenumii}{(C\arabic{enumi}\alph{enumii})}

\begin{enumerate}
\item $S$ is infinite, locally finite: for all $s\in S$ and $r>0$, $%
|\mathcal{B}(s,r)\cap S|=O(r^{d})$, with $\mathcal{B}(s,r)$ $d$-dimensional ball  of center $s$ and radius $r$;
 ${D}_n=\{s_1,\ldots, s_n\}$,  $n \ge 1 $, is a sequence of
arbitrary subsets of $S$ such that $d_n=|D_n|\rightarrow \infty$ as $n\rightarrow\infty$;
\item\label{a:mixing} $Z$ is a   Gaussian random field with covariance function $C(h;\theta)$.
It is also $\alpha$-mixing with  mixing coefficient
$\alpha (m)=\alpha(\infty,\infty,m)$   satisfying:
\begin{enumerate}
\item $\exists\,\eta>0$ s.t. $\sum_{s_k,s_l\in D_{n}}\alpha(\delta(s_k,s_l))^{%
\frac {\eta}{2+\eta}}=O(d_{n})$,
\item $\sum_{m\geq0}m^{d-1}\alpha(m)<\infty$;
\end{enumerate}
\item $\theta^*$,
the true unknown value of the parameter, is an interior point of a compact set $\Theta$ of $\mathbb{R}^p$;
\item  the function  $\theta \mapsto C(h;\theta)$
has continuous second order partial derivatives    with respect to $\theta\in\Theta$,
and these functions are continuous with respect to $h$.
and $\inf_{ \theta\in
\Theta}C(h;\theta)> 0$;
\item\label{a:cl}
the composite likelihood estimator is given by
$
\widehat{\theta}_n=\operatorname{argmin}_{\theta \in \Theta}Q_n(\theta)
$,
where
\begin{equation}
Q_n(\theta)=\frac{1}{d_n}\sum_{k\in D_n}g_k(Y(k);\theta),\qquad \theta\in \Theta,
\end{equation}
with $Y(k)=(Z(s), s\in V_k)'$ and $|V_k|\le B$ for all $k\in S$;

\item the function $g_k(Y(k);\theta)$ is defined as
  $$g_k(Y(k);\theta)=-(1/2)\sum_{l\in V_k, l\ne k}l_{(k,l)}(\theta)$$ where  $l_{(k,l)}=l_{kl},l_{k|l},l_{k-l}$,
  The functions $l_{(k,l)}$ are defined as in (\ref{eq:sub-lik-m}),(\ref{eq:sub-lik-c}) and (\ref{eq:sub-lik-d});
\item\label{a:ident} the function
$\overline{Q}_n(\theta)=\E_{\theta^*}[Q_n(\theta)]$
has a unique global minimum over $\Theta$ at $\theta^*$.
\end{enumerate}
\renewcommand{\labelenumi}{\arabic{enumi}.}
Remarks
\begin{enumerate}
\item The assumption C\ref{a:mixing} is satisfied for  a stationary Gaussian random field
on regular lattice with correlation
function $C(h;\theta)=O(\|h\|)^{-c}$, for some $c> d$
  and its spectral density bounded below \citep[][Corollary 2, p. 59]{Doukhan:1994}.
  \item The assumption C\ref{a:ident}  is an identifiability condition. For each $s$, the function
$\E_{\theta^*}[g_s(Y_s;\theta) ]$ has a global
minimum at $\theta^*$ according   the  Kullback-Leibler inequality
 but in the multidimensional case ($p >1$) $\theta^*$
fails, in general, to be the unique minimizer.
\item The assumption C\ref{a:cl} is satified if we suppose a cut-off weight function for $w_{kl}$.
\item Any individual log-likelihood $l_{(i,j)}$  can be written   as
$$
l_{(k,l)}=c_{1}(\theta,k-l)+c_{2}(\theta, k-l)Z_k^2+c_{3}(\theta,k-l)Z_k^2+c_{4}(\theta,k-l)Z_k Z_l,
$$
where the functions $c_{i}$, $i=1,\ldots,4$ are $\mathcal{C}^{2}$ functions with respect to $\theta$.
\end{enumerate}

\renewcommand{\labelenumi}{\arabic{enumi}:}
\subsection*{Consistency}
Given the previous assumptions C1-C\ref{a:ident}, $\widehat{\theta}_n$ is a consistent estimator for $\theta_0$
provided that $\sup_{\theta\in \Theta}|Q_n(\theta)-\overline{Q}_n(\theta)|\rightarrow 0$
in  probability, as $n\rightarrow\infty$. According Corollary 2.2 in \citet{Newey:1991}, we have to prove
that
\begin{enumerate}
  \item
for each $\theta\in\Theta$, $Q_n(\theta)-\overline{Q}_n(\theta)\rightarrow 0$
in  probability, as $n\rightarrow\infty$;
\item for  $M_n=O_p(1)$,
$$|\overline{Q}_n(\theta') - \overline{Q}_n(\theta)|\le M_n \|\theta' -\theta\|.$$
\end{enumerate}
We sketch the proof for $l_{(k,l)}=l_{kl}$, the same arguments apply for the other sub-likelihoods, using the fourth remark.

\begin{enumerate}
\item We prove that $\sup_{k\in D_n}\E[(\sup_{\theta\in\Theta} g_k(Y(k);\theta))^{2+\eta}]<\infty$, for $\eta >0$.
In fact, we have
\begin{eqnarray*}
g_k(Y(k);\theta)&=&\frac{1}{2}\sum_{l\in V_k, l\ne k}\left\{2\log \sigma^2+\log(1-\rho^2_{kl})+\frac {Z_k^2+Z_l^2-2
 \rho_{kl} Z_kZ_l} {\sigma^2(1-\rho^2_{kl} )}\right\}\\
 &\le&
 \sum_{l\in V_k, l\ne k}\log \sigma^2+\frac12\log(1-\rho^2_{kl})+\frac {Z_k^2+Z_l^2} {\sigma^2(1-\rho^2_{kl} )}  \\
 &\le&
  c_1 |V_k|\log \sigma^2+c_2|V_k|Z_k^2 +c_2\sum_{l\in V_k, l\ne k} Z_l^2
\end{eqnarray*}
and $|V_k|  $ is uniformly bounded according the assumption C\ref{a:cl}.
The uniform bounded moments  $g_k(Y(k);\theta)$ entail uniform $L^1$ integrability of $g_k$ and with the  assumption
C\ref{a:mixing}  we obtain
 \citep[][Theorem 3]{Jenish:Prucha:2009}
$$
Q_n(\theta)-\overline{Q}_n(\theta)=d_n^{-1}\sum_{k\in D_n} \left\{g_k(Y(k),\theta) - \E_\theta[g_k(Y(k),\theta)]\right\}\rightarrow 0, \textrm{ in  probability}
$$

\item We have
\begin{eqnarray*}
  |g_k(Y(k);\theta')-g_k(Y(k);\theta) |&=&\frac{1}{2}\sum_{l\in V_k, l\ne k}
  \left|
  2\log \frac{\sigma^{'2}}{\sigma^{2}}+\log\frac{1-\rho^{'2}_{kl}}{1-\rho^{2}_{kl}}
  \right.\\
&& +(Z^2_k+Z^2_l)\left[\frac {1} {\sigma^{'2}(1-\rho^{'2}_{kl} )}-\frac {1} {\sigma^2(1-\rho^2_{kl} )}\right]\\
  & & -  \left.
  2Z_kZ_l\left[\frac {\rho^{'}_{kl} } {\sigma^{'2}(1-\rho^{'2}_{kl} )}-\frac {\rho_{kl} } {\sigma^2(1-\rho^2_{kl} )}\right]
  \right|\\
  &\le& c_1 |V_k|\|\theta'- \theta\|+c_2 (|V_k| Z_k^2 +\sum_{l\in V_k, l\ne k} Z_l^2)\|\theta'- \theta\|
\end{eqnarray*}
\begin{eqnarray*}
  |\overline{Q}_n(\theta') - \overline{Q}_n(\theta)|
&\le&d_n^{-1}\sum_{k\in D_n}| q_k(\theta') - q_k(\theta) |
\\
&\le&c_3 d_n^{-1}\sum_{k\in D_n}(1+Z_k^2+\sum_{l\in V_k, l\ne k}Z_l^2)\|\theta'- \theta\|
\\
&=&M_n\|\theta'- \theta\|
\end{eqnarray*}
for some positive constants $c_1$, $c_2$ and $c_3$ and
$M_n=c_3 d_n^{-1}\sum_{k\in D_n}(1+Z_k^2+\sum_{l\in V_k, l\ne k}Z_l^2)$.

Since $\E_\theta[M_n]<\infty$, we obtain  the desired result.
\end{enumerate}

\renewcommand{\labelenumi}{N\arabic{enumi}:}
\subsection*{Asymptotic normality}
We make the additional assumption:
\begin{enumerate}
\item there exists two symmetric positive definite matrices $I$ and $J$ such that for
large $n$:

$J_{n}=\var_\theta(\sqrt{d_{n}}\nabla Q_{n}(\theta ))\geq J>0$\,\, and \,\,$I_{n}=\E_{\theta
}(\nabla^2Q_{n}(\theta ))\geq I>0$.
\end{enumerate}
We note that because $g_{s}$ is a
$\mathcal{C}^{2}$ and $\Theta$ is a compact space there exists a   random variable $h(Y(s))$, $\E_{\theta}(h(Y(s)))<\infty$ satisfying:\vspace*{2pt}%
\begin{equation*}
\left|\frac{\partial^2}{\partial \theta_k\theta_l}  g_{s}(Y(s),\theta)\right|^2 \leq h(Y(s)).\vspace*{2pt}
\end{equation*}%
Moreover for all $s\in S$,  $\E_{\theta}[\frac{\partial}{\partial \theta_k} g_{s}(\theta)]=0$,
because $g_s$ is a sum of  log-likelihoods, and it is easy to show that  we have that
$\sup_{s\in S,\,\theta \in \Theta}\E_\theta\left[\left|
\frac{\partial}{\partial \theta_k}
 g_{s}(\theta )\right|^{2+\eta}\right]<\infty$ and
 $\sup_{s\in S,\,\theta \in \Theta}\E_\theta\left[\left|
\frac{\partial^2}{\partial \theta_k\theta_l}
 g_{s}(\theta )\right|^{2+\eta}\right]<\infty .$

 Under the condition C1-C\ref{a:ident} and N1, conditions   (H1-H2-H3)
 of Theorem 3.4.5 in  \citet{Guyon:1995} are satisfied and  $$\sqrt{d_n}J_n^{-1/2}I_n
(\hat\theta_n-\theta^*) \overset{{d}}{\rightarrow }\mathcal{N}({0},\mathcal{I}_p).$$

\bibliographystyle{ECA_jasa}
\bibliography{mybib}


\end{document}